\documentclass[nohyper]{JHEP3} 
\usepackage{epsfig}
\usepackage{cite}
\usepackage{amssymb}
\usepackage{rotating}
\newcommand{\bc}{\begin{center}}
\newcommand{\ec}{\end{center}}
\newcommand{\be}{\begin{equation}}
\newcommand{\ee}{\end{equation}}
\newcommand\fverb{\setbox\pippobox=\hbox\bgroup\verb}
\newcommand\fverbdo{\egroup\medskip\noindent%
            \fbox{\unhbox\pippobox}\ }
\newcommand\fverbit{\egroup\item[\fbox{\unhbox\pippobox}]}
\newbox\pippobox
\title{Towards understanding $b\overline{b}$ production in $\gamma\gamma$
collisions}

\author{by J. Ch\'{y}la
\\
    Center of Particle Physics,
    Institute of Physics AS CR, Na Slovance 2, Prague 8, Czech Republic\\
    E-mail: \email{chyla@fzu.cz}}

\abstract{
Understanding the data on the total cross section
$\sigma_{tot}($e$^+$e$^-\rightarrow$e$^+$e$^-b\overline{b})$ measured at
LEP2 represents a serious challenge for perturbative QCD. In order
to unravel the origins of the discrepancy between data and theory, we
investigate the dependence of four contributions to this cross section on
$\gamma\gamma$ collision energy. As the reliability
of the existing calculations of
$\sigma_{tot}($e$^+$e$^-\rightarrow$e$^+$e$^-b\overline{b})$ depends, among
other things, on the stability of calculations of the cross
section $\sigma_{tot}(\gamma\gamma\rightarrow b\overline{b})$ with respect to
variations of the renormalization and factorization scales, we
investigate this aspect in detail. We show that in most of
the region relevant for the LEP2 data the existing QCD
calculations of $\sigma_{tot}(\gamma\gamma\rightarrow b\overline{b})$
do not exhibit a region of local stability and should thus be taken with
caution. The source of this instability is suggested and its phenomenological
implications for LEP2 data are discussed.}

\begin{document}

\section{Introduction}
\label{intro}
Heavy quark production in hard collisions of hadrons, leptons and
photons has been considered as a clean test of perturbative
QCD. It has therefore come as a surprise that the first data on the
$\overline{b}b$ production in $\overline{\mathrm{p}}$p collisions at the
Tevatron \cite{d0b,cdfb}, $\gamma$p collisions at HERA \cite{h1b,zeusb}
and $\gamma\gamma$ collisions at LEP2 \cite{l3b,opalb} have turned out to
lie significantly and systematically above theoretical
calculations. The disagreement between data \cite{l3b,opalb} and theory
\cite{zerwas,kramer,laenen} was particularly puzzling for the
collisions of two quasireal photons at LEP2.

\begin{figure}\centering\unitlength=1mm
\begin{picture}(160,80)
\put(83,7){
           \begin{sideways}
           \begin{sideways}
           \begin{sideways}
           \epsfig{file=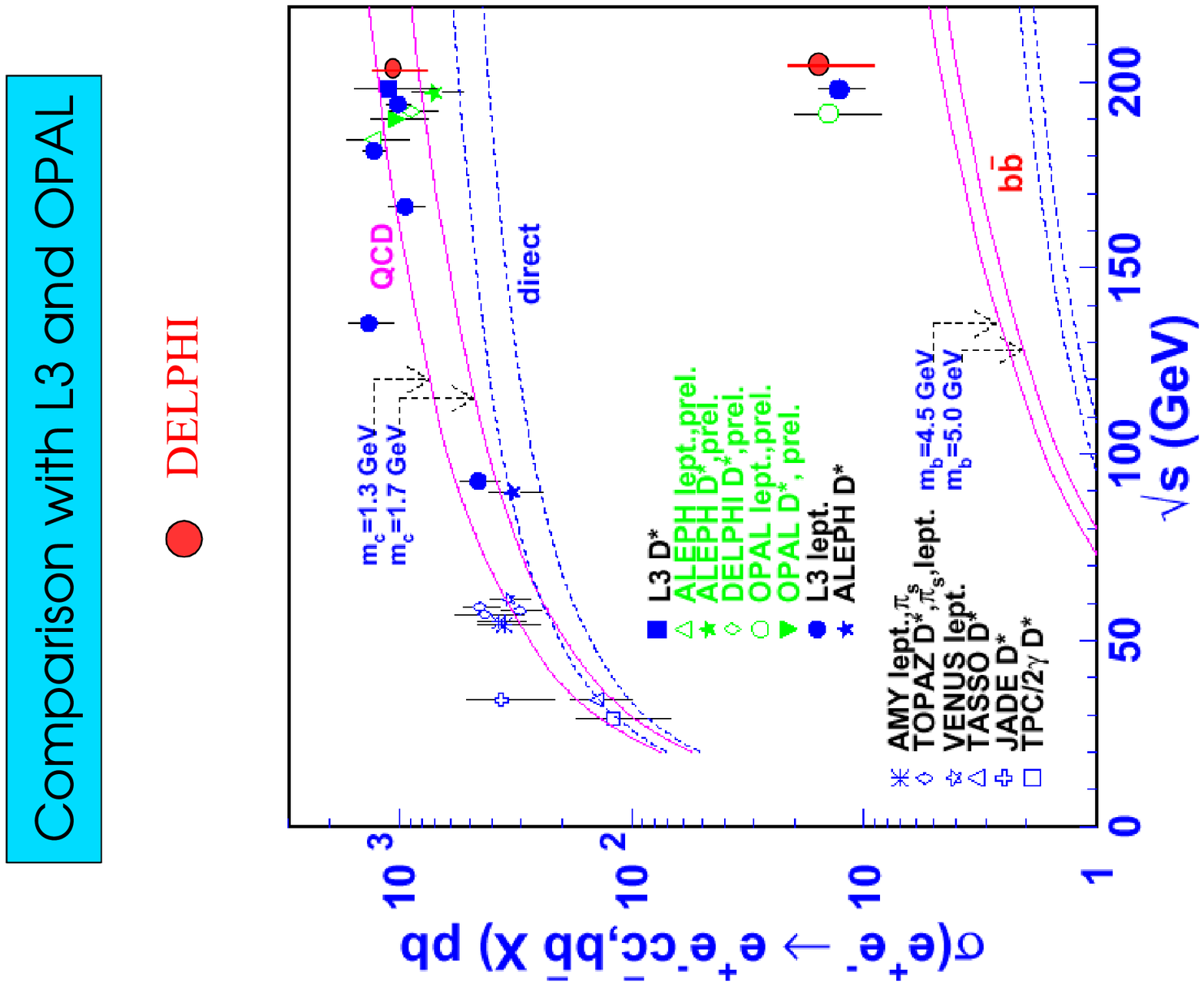,width=7.5cm}
           \end{sideways}
           \end{sideways}
           \end{sideways}
           }
\put(-10,-5){\epsfig{file=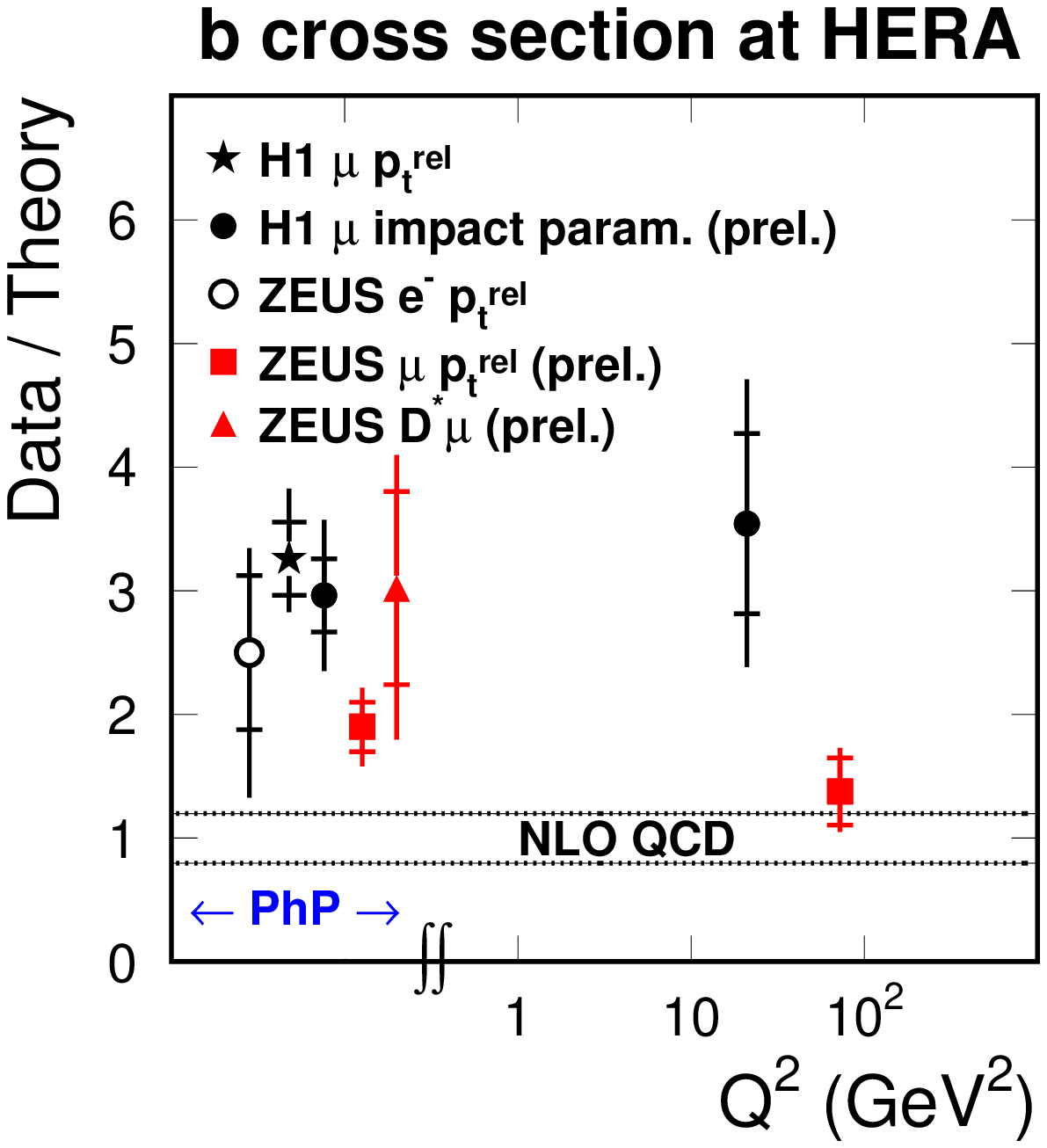,width=10cm}}
\end{picture}
\caption{The current situation with data on $\overline{b}b$
production in ep and e$^+$e$^-$ collisions, including the
most recent data of ZEUS \cite{HERAnew} and DELPHI \cite{DELPHI}.}
\label{bdata}
\end{figure}

The arrival of new data on $\overline{b}b$ production in ep collisions
at HERA \cite{HERAnew}, shown in the left part of Fig. \ref{bdata} as
solid squares, have further complicated the situation. In the range of
moderate $Q^2\simeq 80$ GeV$^2$ the new ZEUS data \cite{HERAnew} are in
reasonable agreement with NLO QCD predictions and also in the
photoproduction region the excess of the new data over theory
is substantially smaller then that in the older data. As a result, there
is now an inconsistency between new ZEUS and older H1 results
\cite{h1b} for moderate $Q^2$, but the situation remains unclear also in
the photoproduction region.
For $\overline{\mathrm{p}}$p collisions the progress on the theoretical
side \cite{ppbarnew,ja} has significantly reduced the discrepancy observed
at the Tevatron.

On the other hand, the problem of understanding the $\overline{b}b$
production in $\gamma\gamma$ collisions remains.
The preliminary DELPHI data presented this spring at PHOTON 2003
conference \cite{DELPHI} and reproduced in Fig. \ref{bdata}, are in
striking agreement with the older L3 and OPAL data. The central values
of all three experiments are almost identical which strongly supports
the reliability of these measurements.
Contrary to the case of analogous discrepancy in antiproton-proton
collisions at the Tevatron, there have been few suggestions how to
explain the sizable excess of data over current theory in
$\gamma\gamma$ collisions. Neither the use of unintegrated parton
distribution functions \cite{Hannes}, nor the production of
supersymmetric particles \cite{Berger}, proposed for explaining
an analogous excess in antiproton-proton collisions, are of much
help for LEP2 data, primarily because of low
$\gamma\gamma$ energies involved. Quite recently,
however, this discrepancy has been interpreted as an evidence for
integer quark charges \cite{integer}. We will come back to this
suggestion in Section 4.1.

In \cite{ja} we have investigated the sensitivity of QCD calculations of
$\sigma_{tot}(p\overline{p}\rightarrow b\overline{b};M,\mu)$ to the
variation of the renormalization and factorization scales $\mu$ and $M$.
In particular we have argued that in order to arrive at locally stable
results \cite{pms} these two scales must be kept independent.
We have furthermore shown that in the Tevatron energy range the position of
the saddle point of the cross section
$\sigma_{tot}(p\overline{p}\rightarrow b\overline{b};S,M,\mu)$ lies far away
from the ``diagonal'' $\mu=M$
used in all existing calculations. Using the NLO prediction at the
saddle point instead of the conventional choice $\mu=M=m_b$ enhances the
theoretical prediction in the Tevatron energy range by a factor of about 2,
which may help explaining the excess of data over NLO QCD predictions.

In this paper similar analysis is performed for $\gamma\gamma$ collisions
at the total centre of mass energies $W$ relevant for existing LEP2 data.
The specific features of the theoretical description of $Q\overline{Q}$
production in $\gamma\gamma$ collisions have been discussed in
\cite{ascona}. However, as all three experiments at LEP2 have measured
merely an integral over the cross section
$\sigma(\gamma\gamma\rightarrow b\overline{b},W)$
weighted by the product of photon fluxes inside the beam electrons and
positrons, is is also important to understand the $W$-dependence
of the four individual contributions to it.

The paper is organized as follows. The basic facts and formulae relevant
for the quantitative investigation of renormalization and factorization
scale dependence of finite order QCD approximations are collected in
Section 2. This is followed in Section 3 by the discussion of the general
form of $\sigma_{tot}(\gamma\gamma\rightarrow b\overline{b};W,M,\mu)$.
In Section 4 the $W$-dependence of the four contributions
to the cross section
$\sigma_{tot}($e$^+$e$^-\rightarrow$e$^+$e$^-b\overline{b})$
is investigated at the LO of QCD. The quantitative role of NLO
corrections and the implications of the (in)stability of existing
calculations of
$\sigma_{tot}^{\mathrm{NLO}}(\gamma\gamma\rightarrow b\overline{b})$
for explaining the observed puzzle is discussed in Section 5.
The conclusions are drawn in Section 6.

\section{Basic facts and formulae}
\label{basics}
The basic quantity of perturbative QCD calculations, the renormalized
color coupling $\alpha_s(\mu)$, depends on the renormalization scale $\mu$
in   a way governed by the equation
\begin{equation}
\frac{{\mathrm d}\alpha_s(\mu)}{{\mathrm d}\ln \mu^2}\equiv
\beta(\alpha_s(\mu))=
-\frac{\beta_0}{4\pi}\alpha_s^2(\mu)-
\frac{\beta_1}{16\pi^2}
\alpha_s^3(\mu)+\cdots,
\label{RG}
\end{equation}
where for $n_f$ massless quarks $\beta_0=11-2n_f/3$ and
$\beta_1=102-38n_f/3$. Its solutions depend beside $\mu$ also on the
renormalization scheme (RS). At the NLO this RS can be specified via the
parameter
$\Lambda_{\mathrm{RS}}$ corresponding to the renormalization scale for
which $\alpha_s$ diverges. The coupling $\alpha_s(\mu)$ then solves the
equation
\begin{equation}
\frac{\beta_0}{4\pi}\ln\left(\frac{\mu^2}{\Lambda^2_{\mathrm{RS}}}\right)=
\frac{1}{\alpha_s(\mu)}+
c\ln\frac{c\alpha_s(\mu)}{1+c\alpha_s(\mu)},~~c\equiv \beta_1/(4\pi\beta_0).
\label{equation}
\end{equation}
At the NLO the coupling $\alpha_s$ is
a function of the ratio $\mu/\Lambda_{\mathrm{RS}}$ and the variation of
the RS for fixed scale $\mu$ is therefore equivalent to the variation of
$\mu$ for fixed RS. To vary both the renormalization scale and scheme is
legitimate, but redundant. Throughout the paper I will work in the
conventional $\overline{\mathrm{MS}}$ RS and vary the renormalization
scale $\mu$ only. As we shall investigate the QCD predictions down to
quite small values of the renormalization scale $\mu$, the equation
(\ref{equation}) will be solved numerically, rather than expanding its solution
in inverse powers of $\ln(\mu/\Lambda)$.

The main difference between hard collisions of hadrons and photons
comes from the fact that quark and gluon distribution functions of the photon
\begin{equation}
\Sigma(x,M) \equiv
\sum_{i=1}^{n_f} \left(q_i(x,M)+\overline{q}_i(x,M)\right),~
q_{\mathrm{NS}}(x,M) \equiv
\sum_{i=1}^{n_f}\left(e_i^2-\langle e^2\rangle\right)
\left(q_i(x,M)+\overline{q}_i(x,M)\right)
\label{definice}
\end{equation}
satisfy the system of coupled inhomogeneous evolution equations
\begin{eqnarray}
\frac{{\mathrm d}\Sigma(M)}{{\mathrm d}\ln M^2}& =&
\delta_{\Sigma}k_q(M)+P_{qq}(M)\otimes \Sigma(M)+ P_{qG}(M)\otimes G(M),
\label{Sigmaevolution}
\\ \frac{{\mathrm d}G(M)}{{\mathrm d}\ln M^2} & =& k_G(M)+
P_{Gq}(M)\otimes \Sigma(M)+ P_{GG}(M)\otimes G(M),
\label{Gevolution} \\
\frac{{\mathrm d}q_{\mathrm {NS}}(M)}{{\mathrm d}\ln M^2}& =&
\delta_{\mathrm {NS}} k_q(M)+P_{\mathrm {NS}}(M)\otimes q_{\mathrm{NS}}(M),
\label{NSevolution}
\end{eqnarray}
where $\delta_{\mathrm{NS}}\equiv 6n_f\left(\langle e^4\rangle-\langle
e^2\rangle ^2\right)$, $\delta_{\Sigma}=6n_f\langle e^2\rangle$ and
\begin{eqnarray}
k_q(x,M) & = & \frac{\alpha}{2\pi}\left[k^{(0)}_q(x)+
\frac{\alpha_s(M)}{2\pi}k_q^{(1)}(x)+
\left(\frac{\alpha_s(M)}{2\pi}\right)^2k^{(2)}_q(x)+\cdots\right],
\label{splitquark} \\ k_G(x,M) & = &
\frac{\alpha}{2\pi}\left[~~~~~~~~~~~~
\frac{\alpha_s(M)}{2\pi}k_G^{(1)}(x)+
\left(\frac{\alpha_s(M)}{2\pi}\right)^2k^{(2)}_G(x)+\cdots\;\right],
\label{splitgluon} \\ P_{ij}(x,M) & = &
~~~~~~~~~~~~~~~~~~\frac{\alpha_s(M)}{2\pi}P^{(0)}_{ij}(x) +
\left(\frac{\alpha_s(M)}{2\pi}\right)^2 P_{ij}^{(1)}(x)+\cdots.
\label{splitpij}
\end{eqnarray}
The lowest order inhomogeneous splitting function
$k_q^{(0)}(x)=(x^2+(1-x)^2)$ as well as the homogeneous
splitting functions $P^{(0)}_{ij}(x)$ are {\em unique},
whereas all higher order splitting functions
$k^{(j)}_q,k^{(j)}_G,P^{(j)}_{kl},j\ge 1$ depend  on the choice of
the factorization scheme (FS). Although potentially important, I
will not exploit this freedom and throughout this paper will stay
within the $\overline{\mathrm{MS}}$ FS. The equations
(\ref{Sigmaevolution}-\ref{NSevolution}) can be recast into evolution
equations for $q_i(x,M),\overline{q}_i(x,M)$ and $G(x,M)$ with
inhomogeneous splitting functions $k_{q_i}^{(0)}=3e_i^2k_q^{(0)}$.

Due to the presence of the inhomogeneous terms on the r.h.s. of
(\ref{Sigmaevolution}-\ref{NSevolution}) their general
solutions can be written as a sum of a particular solution of the full
inhomogeneous equations and a general solution, called {\em
hadron-like} (HAD), of the corresponding homogeneous ones. A
subset of the former resulting from the resummation of
contributions of diagrams in Fig. \ref{pointlike} describing multiple
parton emissions off
the primary QED vertex $\gamma\rightarrow q\overline{q}$ and
vanishing at $M=M_0$, are called {\em point-like} (PL) solutions.
\FIGURE{
\epsfig{file=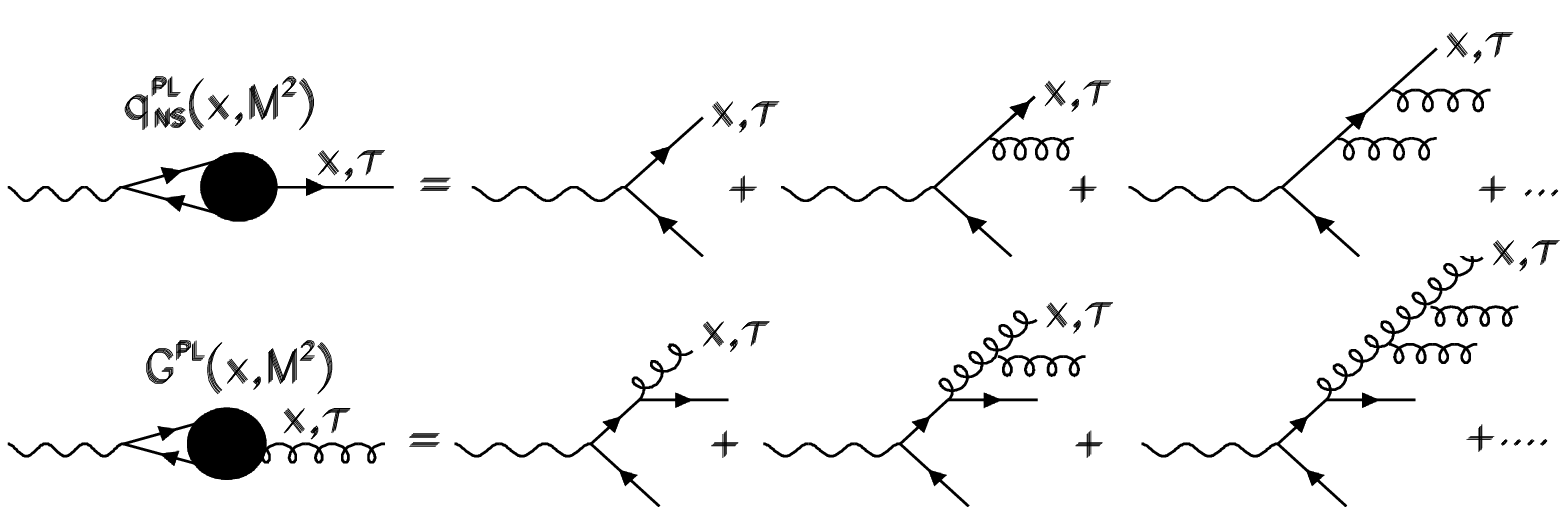,width=14cm}
\caption{Diagrams defining the pointlike parts of the
nonsinglet quark and gluon distribution functions of the photon,
represented by the symbols on the left.}
\label{pointlike}
}
Due to the arbitrariness in the choice of $M_0$ the separation
\begin{equation}
D(x,M)= D^{\mathrm {PL}}(x,M,M_0)+D^{\mathrm{HAD}}(x,M,M_0).
\label{separation}
\end{equation}
is, however, ambiguous. The explicit form of the pointlike contribution
to nonsinglet distribution function
\begin{equation}
q_{\mathrm {NS}}^{\mathrm {PL}}(n,M_0,M)=\frac{4\pi}{\alpha_s(M)}
\left[1-\left(\frac{\alpha_s(M)}{\alpha_s(M_0)}\right)^
{1-2P^{(0)}_{qq}(n)/\beta_0}\right]a_{\mathrm {NS}}(n),
\label{explicit}
\end{equation}
where
\begin{equation}
a_{\mathrm {NS}}(n)\equiv \frac{\alpha}{2\pi\beta_0}
\frac{k_{\mathrm {NS}}^{(0)}(n)}{1-2P^{(0)}_{qq}(n)/\beta_0}
\label{ans}
\end{equation}
is often claimed to show that it behaves as ${\cal O}(\alpha/\alpha_s)$.
However, the fact that $\alpha_s(M)$ appears in the denominator of
$q^{\mathrm {PL}}_{\mathrm {NS}}$ cannot be interpreted in this way
\cite{factor}
because switching QCD off by sending $\Lambda_{\mathrm{RS}}\rightarrow 0$
for fixed $M,M_0$ reduces, as expected, the expression (\ref{explicit})
to the purely QED contribution, corresponding to the first diagram in
in the upper part of Fig. \ref{pointlike}
\begin{equation}
{ q^{\mathrm {PL}}_{\mathrm {NS}}(x,M,M_0)\rightarrow}
{\frac{\alpha}{2\pi} k_{\mathrm {NS}}^{(0)}(x)\ln\frac{M^2}{M_0^2}}.
\label{PL}
\end{equation}
The form (\ref{explicit}) merely implies that for asymptotically large
$M$ the pointlike part $q^{\mathrm {PL}}_{\mathrm {NS}}$ behaves as $\ln M$.

As emphasized long time ago by Politzer \cite{politzer} there is no compelling
reason for identifying the
renormalization and factorization scales $\mu$ and $M$ and one
should therefore keep these scale as independent free parameters of
any finite order perturbative calculations.

\section{$Q\overline{Q}$ production in $\gamma\gamma$ collisions}
\label{overall}
We shall first recall the general form of QCD expression for
$\sigma_{tot}(\gamma\gamma\rightarrow Q\overline{Q})$
and then discuss in detail the renormalization and
factorization scale dependence of its finite order approximations.

\subsection{General form of
$\sigma_{tot}(\gamma\gamma\rightarrow Q\overline{Q})$}
\label{general}
In the calculations of refs. \cite{zerwas,kramer,laenen},
performed with fixed pole quark masses, the NLO QCD
approximation to $\sigma_{tot}(\gamma\gamma\rightarrow Q\overline{Q})$
is defined by taking into account the first two terms in the expansions
of direct, as well as single and double resolved photon contributions
\begin{eqnarray}
\sigma_{\mathrm{dir}}(M) &= &
\sigma_{\mathrm{dir}}^{(0)}+
\sigma_{\mathrm{dir}}^{(1)}\alpha_s(\mu)+
\sigma_{\mathrm{dir}}^{(2)}(M,\mu)\alpha_s^2(\mu)+
\sigma_{\mathrm{dir}}^{(3)}(M,\mu)\alpha_s^3(\mu)+\cdots,
~\label{dir}\\
\sigma_{\mathrm{sr}}(M) & = &
~~~\sigma_{\mathrm{sr}}^{(1)}(M)\alpha_s(\mu)+
\sigma_{\mathrm{sr}}^{(2)}(M,\mu)\alpha_s^2(\mu)+
\sigma_{\mathrm{sr}}^{(3)}(M,\mu)\alpha_s^3(\mu)+\cdots,
\label{sr}\\
\sigma_{\mathrm{dr}}(M) & =&
~~~~~~~~~~~~~~~~~~~~~~~~~~~
\sigma_{\mathrm{dr}}^{(2)}(M)\alpha_s^2(\mu)+
\sigma_{\mathrm{dr}}^{(3)}(M,\mu)\alpha_s^3(\mu)+\cdots.
\label{dr}
\end{eqnarray}
Starting at order $\alpha_s^2$ the direct
photon contribution depends also on the factorization scale and therefore
mixes with the single and double resolved photon ones. The first two terms
in (\ref{dir}) are, however, totally unrelated to any terms in
(\ref{sr}) or (\ref{dr}).

The approximations employed in \cite{zerwas,kramer,laenen} include all
terms that are currently known, so we cannot do better at this moment. On
the other hand we should be aware of its theoretical deficiency. The fact
that the first two terms of (\ref{dir}-\ref{dr})
start and end at different powers of $\alpha_s$ is usually justified by
claiming that PDF of the photon, which appear in expressions for
$\sigma_{\mathrm{sr}}^{(1,2)}(M)$ and
$\sigma_{\mathrm{dr}}^{(2,3)}(M)$, behave as $\alpha/\alpha_s$.
Consequently, the first terms in all three expressions
(\ref{dir}-\ref{dr}) are claimed to be of order $(\alpha_s)^0=1$ and
the second ones of order $\alpha_s$.
However, as emphasized above and argued
in detail in \cite{factor}, the term $\ln M^2$ characterizing the large
$M$ behaviour of PDF of the photon comes from integration over the
transverse degree of freedom of the purely QED vertex
$\gamma\rightarrow q\overline{q}$ and cannot therefore be interpreted as
$1/\alpha_s(M)$.

\subsection{Direct photon contribution}
\label{dircont}
For proper treatment of the direct photon contribution (\ref{dir})
to $\sigma_{tot}(\gamma\gamma\rightarrow Q\overline{Q})$,
the total cross section
of e$^+$e$^-$ annihilations into hadrons at center-of-mass energy
$\sqrt{S}$ provides a suitable guidance. For $n_f$ massless quarks
we have
\begin{equation}
\sigma_{\mathrm{had}}(\sqrt{S})=
             \sigma_{\mathrm{had}}^{(0)}(\sqrt{S})+
\alpha_s(\mu)\sigma_{\mathrm{had}}^{(1)}(\sqrt{S})+
\alpha_s^2(\mu)\sigma_{\mathrm{had}}^{(2)}(\sqrt{S}/\mu)+\cdots=
\sigma_{\mathrm{had}}^{(0)}(1+r(\sqrt{S})),
\label{Rlarge}
\end{equation}
where the term $\sigma_{\mathrm{had}}^{(0)}(\sqrt{S})\equiv
(4\pi\alpha^2/3S)\sum_{f=1}^{n_f}e_f^2$ comes,
similarly as $\sigma_{\mathrm{dir}}^{(0)}$ in (\ref{dir}), from
pure QED, whereas genuine QCD effects are contained in the quantity
$r(\sqrt{S})$
\begin{equation}
r(\sqrt{S})=\frac{\alpha_s(\mu)}{\pi}
\left[1+\alpha_s(\mu)r_1(\sqrt{S}/\mu)+\cdots\right].
\label{rsmall}
\end{equation}
For the purpose of QCD analysis of the quantity (\ref{Rlarge}) it
is a generally accepted practice to discard  the lowest order term
$\sigma_{\mathrm{had}}^{(0)}$
and denote as the ``leading order'' the second term in (\ref{Rlarge}),
i.e. $\sigma_{\mathrm{had}}^{(0)}\alpha_s(\mu)/\pi$. The adjectives
``LO'' and ``NLO'' are thus reserved for genuine QCD effects
described by $r(\sqrt{S})$. The rationale for this terminology is simple:
to work in a well-defined renormalization scheme of $\alpha_s$ requires
including in (\ref{rsmall}) at least first two consecutive powers of
$\alpha_s(\mu)$. The explicit $\mu$-dependence of
$r_1(\sqrt{S}/\mu)$ cancels to the order $\alpha_s^2$ the
implicit $\mu$-dependence of the leading order term
$\alpha_s(\mu)/\pi$ in (\ref{rsmall}) and thus guarantees that the
derivative with respect to $\ln \mu$ of the sum of first two terms in
(\ref{rsmall}) behaves as $\alpha_s^3$. For purely perturbative
quantities like (\ref{Rlarge}) the
association of the term ``NLO QCD approximation'' with a well-defined
renormalization scheme is a generally accepted convention, worth
retaining for any physical quantity, like the direct photon contribution
$\sigma_{\mathrm{dir}}$ in (\ref{dir}).
\FIGURE{
\epsfig{file=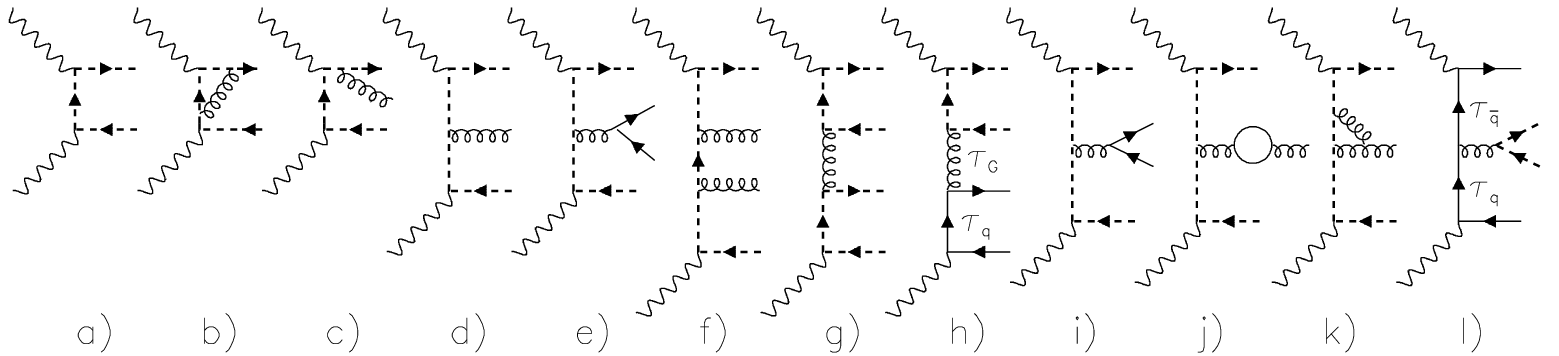,width=\textwidth}
\caption{Examples of diagrams describing the direct photon contribution
to $\sigma(\gamma\gamma\rightarrow Q\overline{Q})$ up to the order
$\alpha^2\alpha_s^2$. The solid (dashed) lines denote light (heavy)
quarks.}
\label{figdir}}

Contrary to this practice, the calculation in refs.
\cite{zerwas,kramer,laenen} consider the purely QED contribution
\be
\sigma^{(0)}_{\mathrm{dir}}(W)=\sigma_0
\left[\left(1+\frac{4m_b^2}{W^2}-\frac{8m_b^4}{W^4}\right)
\ln\frac{1+\beta}{1-\beta}-\beta\left(1+\frac{4m_b^2}{W^2}\right)\right],
~\sigma_0\equiv \frac{12\pi e_b^4\alpha^2}{W^2},
\label{pureqed}
\ee
where $\beta=\sqrt{1-4m_b^2/W^2}$, as the LO approximation.
This is legitimate but implies that their NLO approximation, includes only
the lowest order term in $\alpha_s$ and cannot therefore
be associated to a well-defined renormalization scheme of $\alpha_s$
even if the NLO expression for $\alpha_s$ is used therein. For QCD
analysis of $\sigma_{\mathrm{dir}}$ in a well-defined
renormalization scheme the incorporation of the third term in
(\ref{dir}), proportional to $\alpha^2\alpha_s^2$, is indispensable.

At the order $\alpha^2\alpha_s^2$ the diagrams with light quarks
appear and we can distinguish three classes of direct
photon contributions differing by the overall heavy quark charge
factor $CF$:
\begin{description}
\item{\bf Class A:} $CF=e_Q^4$. Comes from diagrams, like those in
Fig. \ref{figdir}e-g, in which both
photons couple to heavy $Q\overline{Q}$ pairs. Despite the presence of
mass singularities in contributions of individual diagrams coming from
gluons and light quarks in the final state and from loops, the KLN
theorem implies that at each order of $\alpha_s$ the sum of all
contributions of this class to $\sigma_{\mathrm{dir}}$ is finite. Note
that the first as well as the second terms in (\ref{dir}) are also
proportional to $e_Q^4$ and it is therefore this class of direct photon
contributions that is needed for the calculation of
$\sigma_{\mathrm{dir}}$ to be performed in a well-defined RS.
\item{\bf Class B:} $CF=e_Q^2$.
Comes from diagrams, like that in Fig. \ref{figdir}h, in which one of
the photons couples to a heavy $Q\overline{Q}$ and the other to a light
$q\overline{q}$ pair. For massless light quarks this diagram has initial
state mass singularity, which is removed by introducing the concept of the
light quark (and gluon) distribution functions of the photon. The
factorization scale dependence of the contribution of this diagram is then
related to that of single resolved photon diagrams in Fig. \ref{srpl}a,c.
\item{\bf Class C:} $CF=1$. Comes from diagrams in which both
photons couple to light $q\overline{q}$ pairs, as those in Fig.
\ref{figdir}l. In this case the analogous subtraction procedure relates
it to the single resolved photon contribution of the diagram in Fig.
\ref{srpl}f and double resolved photon contribution of the
diagram in Fig. \ref{srpl}h. The classes B and C are thus needed to
guarantee
the factorization scale (and scheme) invariance of the single and double
resolved photon contributions to order $\alpha^2\alpha_s^2$.
\end{description}
Because of different charge factors $CF$,
the classes A, B and C do not mix under renormalization of $\alpha_s$ and
factorization of mass singularities. As the diagrams in Fig. \ref{figdir}e
and \ref{figdir}l give the same final state $q\overline{q}Q\overline{Q}$,
we should consider their interference term as well, but it turns out that
it does not contribute to the total cross section
$\sigma(\gamma\gamma\rightarrow Q\overline{Q})$.

\subsection{Resolved photon contribution}
\label{rescont}
\FIGURE{
\epsfig{file=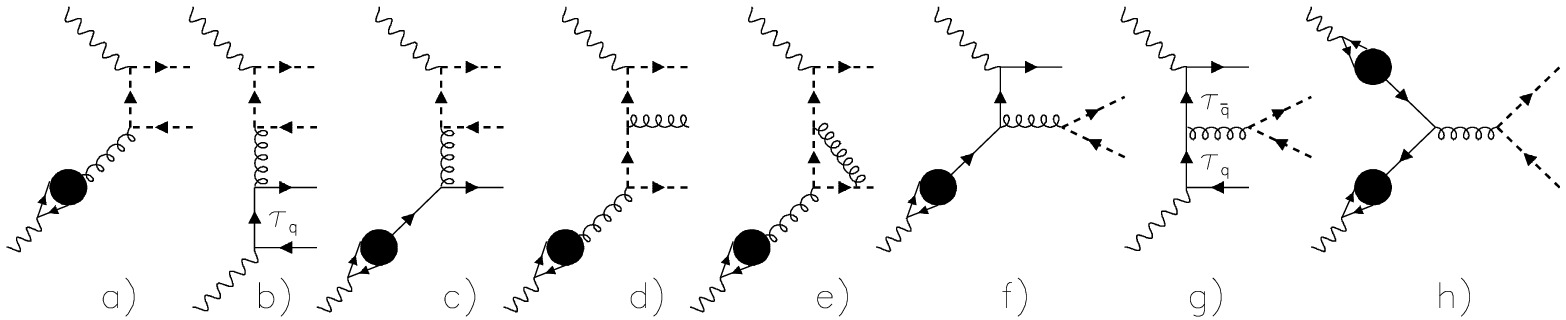,width=\textwidth}
\caption{Examples of resolved photon diagrams involving the pointlike
parts of PDF of the photon and the related direct photon diagrams.}
\label{srpl}}
\noindent
The classes B and C of direct photon contributions
of the order $\alpha^2\alpha_s^2$ are indispensable to
render the sum of direct and resolved photon contributions factorization
scale invariant up to order $\alpha^2\alpha_s^2$. To see this in
detail, let us write the sum of first two terms in (\ref{sr}-\ref{dr})
explicitly in terms of PDF and parton level cross sections
\begin{eqnarray}
\sigma_{\mathrm{res}}^{(12)}(M,\mu)&\equiv&
2\alpha_s(\mu)\int\!\!{\mathrm{d}}xG(x,M)
\left[\sigma_{\gamma G}^{(1)}(x)+\alpha_s(\mu)\sigma_{\gamma G}^{(2)}
(x,M,\mu)\right]+\label{l3}\nonumber \\
& & 4\alpha_s^2(\mu)\int\!\!{\mathrm{d}}x\sum_iq_i(x,M)
\sigma_{\gamma q_i}^{(2)}(x,M)+\label{kk}\nonumber\\
& & 2\alpha_s^2(\mu)\int\!\!\!\int
{\mathrm{d}}x{\mathrm{d}}y\sum_i q_i(x,M)\overline{q}_i(y,M)
\left[\sigma_{q\overline{q}}^{(2)}(xy)+\alpha_s(\mu)
\sigma_{q\overline{q}}^{(3)}(xy,M,\mu)\right]+
\label{p4}\nonumber\\
& &~\alpha_s^2(\mu)\int\!\!\!\int
{\mathrm{d}}x{\mathrm{d}}yG(x,M)G(y,M)
\left[\sigma_{GG}^{(2)}(xy)+\alpha_s(\mu)\sigma_{GG}^{(3)}(xy,M,\mu)
\right]+\label{l33}\nonumber \\
& & 2\alpha_s^3(\mu)\int\!\!\!\int
{\mathrm{d}}x{\mathrm{d}}y\Sigma(x,M)G(y,M)
\sigma_{qG}^{(3)}(xy,M)
\label{resolved}
\end{eqnarray}
where $\sum_iq_i$ runs over $n_f$ quark flavors and the factors of two
and four reflect the identity of beam particles and equality of
contributions from quarks and antiquarks inside the beam photons.
Recalling the general form of the derivative
${\mathrm{d}}\sigma_{\mathrm{res}}/{\mathrm{d}}\ln M^2$
\begin{eqnarray}
\lefteqn{\!\!\!\!\!\!\!
\frac{{\mathrm{d}}\sigma_{\mathrm{res}}}{{\mathrm{d}}\ln M^2}
=\int{\mathrm{d}}x W_0(x,M)+\!\!
\int\!\!{\mathrm{d}}x\!\left[\sum_iq_i(x,M)W_{q_i}(x,M)+
G(x,M)W_G(x,M)\right]+} & &
\label{der}\nonumber \\
& &\!\!\!\!\!\!\!\!
\int\!\!\!\!\int\!{\mathrm{d}}x{\mathrm{d}}y
\!\left[G(x,M)G(y,M)W_{GG}(xy,M)\!+\!\sum_i
q_i(x,M)\overline{q}_i(y,M)W_{q\overline{q}}(xy,M)
+\right. \label{l54}\nonumber \\
& & ~~~~~~~~ \Sigma(x,M)G(y,M)W_{qG}(xy,M) \Bigg],
\label{derivace}
\end{eqnarray}
using (\ref{Sigmaevolution}-\ref{NSevolution}) and denoting
$\alpha_s\equiv\alpha_s(\mu)$,
$\dot{f}\equiv \mathrm{d}f/\mathrm{d}\ln M^2$ we find
\begin{eqnarray}
W_0(x,M)&=&\frac{\alpha\alpha_s^2}{\pi}\left\{
\frac{1}{2\pi}k_G^{(1)}(x)\sigma_{\gamma G}^{(1)}(x)+
6k_q^{(0)}(x)\sum_i e_i^2\sigma_{\gamma q_i}^{(2)}(x,M)\right\}+
\cdots
\label{W0} \\
W_{q_i}(x,M)&=& \frac{\alpha_s^2}{\pi}
\!\left\{4\pi\dot{\sigma}_{\gamma q}^{(2)}(x)+
\!\!\int\!\!{\mathrm{d}}z\!\left[P_{Gq}^{(0)}(z) \sigma_{\gamma
G}^{(1)}(xz)+3e_i^2\alpha k_q^{(0)}(z)
\sigma_{q\overline{q}}^{(2)}(xz)\right]\right\}\!\!+\cdots
\label{Wq}\\
W_G(x,M)&=& \frac{\alpha_s^2}{\pi} \!\left\{2\pi\dot{\sigma}_{\gamma
G}^{(2)}(x)+ \!\!\int\!\!{\mathrm{d}}zP_{GG}^{(0)}(z)
\sigma_{\gamma G}^{(1)}(xz)\right\}+\cdots
\label{WG}\\
W_{GG}(x,M)& = & \frac{\alpha_s^3}{\pi}
\left\{\pi\dot{\sigma}_{GG}^{(3)}(x)+\!\!\int\!\!
{\mathrm{d}}zP^{(0)}_{GG}(z)\sigma_{GG}^{(2)}(xz)\right\}+\cdots
\label{WGG}\\
W_{q\overline{q}}(x,M)& = & \frac{\alpha_s^3}{\pi}
\left\{2\pi\dot{\sigma}_{q\overline{q}}^{(3)}(x)+2\!\!\int\!\!
{\mathrm{d}}zP^{(0)}_{qq}(z)\sigma_{q\overline{q}}^{(2)}(xz)\right\}
+\cdots\label{Wqq}\\
W_{qG}(x,M)& = & \frac{\alpha_s^3}{\pi}
\left\{2\pi\dot{\sigma}_{qG}^{(3)}(x)+\!\!\int\!\!
{\mathrm{d}}z\!\left[P^{(0)}_{qG}(z)\sigma_{q\overline{q}}^{(2)}(xz)+
P^{(0)}_{Gq}(z)\sigma_{GG}^{(2)}(xz)\right]\!\right\}\!+\cdots
\label{WqG}
\end{eqnarray}
Only the lowest order terms on the r.h.s. of (\ref{W0}-\ref{WqG})
are written out explicitly. All integrals in (\ref{Wq}-\ref{WqG})
go formally from $0$ to $1$, but threshold behaviour of cross
sections $\sigma_{ij}(xz)$ restricts the region to $xz\ge 4m_Q^2/W^2$.

The factorization scale invariance of (\ref{resolved}) requires
that its variation with respect to $\ln M^2$ is
of higher order in $\alpha_s$ than the approximation
itself. There is no dispute that direct photon contributions of classes
B and C are needed to guarantee this property. The question is which
terms on the r.h.s. of (\ref{derivace}) must vanish if the
approximation is defined by (\ref{resolved}).

In the conventional approach both $q(M)$ and $G(M)$ are claimed to
be of order $\alpha/\alpha_s$ and the approximation (\ref{resolved})
thus of the order $\alpha^2\alpha_s$, implying that only terms
up to this order must vanish in (\ref{derivace}). This in turn means
that the functions (\ref{Wq}-\ref{WG}) must vanish to order
$\alpha_s^2$ and (\ref{WGG}-\ref{WqG}) to order $\alpha_s^3$
respectively, which, indeed, they do
\footnote{The latter condition is actually the same as for
$Q\overline{Q}$ production in hadron-hadron collisions.}.
The fact that the expression on the r.h.s. of (\ref{W0}) does not
vanish is of no concern in this approach as it is manifestly of
the order $\alpha\alpha_s^2$ and thus supposedly of higher order than
(\ref{resolved}) itself.

If, on the other hand, we take into account that quark and gluon
distribution functions of the photon behave as
$q(M),G(M)\propto\alpha$, we see that $W_0$ is of the same
order $\alpha^2\alpha_s^2$ as the products $q_iW_{q_i}, GW_G$ and
other integrands on the r.h.s. of (\ref{derivace}) and must therefore
also vanish for theoretical consistency of the approximation
(\ref{resolved}). This, in turn, necessitates the inclusion of class
B direct photon contributions of the order $\alpha\alpha_s^2$, like
those in Fig. \ref{srpl}b,g, which provide the $M$-dependent terms
the derivative of which cancels the first term in (\ref{derivace})
involving the integral over $W_0$. Note that $W_{q_i}$
in (\ref{Wq}) receives contributions from the derivatives of both
single and double resolved photon diagrams, proportional to
$\sigma_{\gamma G}$ and $\sigma_{q\overline{q}}$, respectively.
This fact reflects the mixing of single and double resolved photon
contributions, which starts at the order $\alpha^2\alpha_s^2$ and is
due to the presence of the inhomogeneous splitting terms in the
evolution equations (\ref{Sigmaevolution}-\ref{NSevolution}). For
theoretical consistency of the sum of direct and resolved
photon contribution up to the order $\alpha\alpha_s^2$ only the
lowest order double resolved photon contribution needs to be included.

\section{$\overline{b}b$ production at LEP2}
\label{lepresults}
We now turn to the phenomenological analysis of $\overline{b}b$
production at LEP2, where the incoming leptons act as sources of transverse
and longitudinal virtual photons, described by the fluxes
\begin{eqnarray}
f^{\gamma}_{T}(y,Q^2) & = & \frac{\alpha}{2\pi}
\left(\frac{1+(1-y)^2)}{y}\frac{1}{Q^2}-\frac{2m_{\mathrm e}
^2 y}{Q^4}\right),
\label{fluxT} \\
f^{\gamma}_{L}(y,Q^2) & = & \frac{\alpha}{2\pi}
\frac{2(1-y)}{y}\frac{1}{Q^2},
\label{fluxL}
\end{eqnarray}
where $Q^2$ stands for photon virtuality. Although the kinematic region
of the LEP data includes photon virtualities up to moderate $Q^2$, the
cross section of the inclusive process
\be
{\mathrm{e}}^+{\mathrm{e}}^-\rightarrow
{\mathrm{e}}^+{\mathrm{e}}^-
b\overline{b}+~{\mathrm{anything}}
\label{process}
\ee
is dominated by the production of the $b\overline{b}$ pair in the
collision of two quasireal photons with very small $Q^2$, typically
$\langle Q^2\rangle \simeq 0.01$ GeV$^2$. For such small $Q^2$ the
cross sections of hard processes involving longitudinal virtual
photons, which are proportional to $Q^2$, are expected to be
negligible compared to those of transverse virtual photons.
When talking about the production of $b\overline{b}$ in
${\mathrm{e}}^+{\mathrm{e}}^-$ collisions we shall always
mean in association with the ${\mathrm{e}}^+{\mathrm{e}}^-$
pair, but for brevity of notation shall drop this latter
specification, writing
$\sigma_{tot}({\mathrm{e}}^+{\mathrm{e}}^-\rightarrow b\overline{b})$.
instead of
$\sigma_{tot}({\mathrm{e}}^+{\mathrm{e}}^-\rightarrow
{\mathrm{e}}^+{\mathrm{e}}^-b\overline{b})$.

Although the data are available only for cross sections integrated over
the whole phase space, we shall discuss the contributions
${\mathrm{d}}\sigma_k({\mathrm{e}}^+{\mathrm{e}}^-
\rightarrow b\overline{b})/{\mathrm{d}}W$ of individual processes as
functions of $\gamma\gamma$ collision energy $W$. The shapes of
these contributions can altenatively be characterized by the functions
\be
F_k(W)\equiv \int_{2m_b}^{W}
{\mathrm{d}}w \frac{{\mathrm{d}}
\sigma_k({\mathrm{e}}^+{\mathrm{e}}^-
\rightarrow b\overline{b})}{{\mathrm{d}}w},~~
G_k(W)\equiv \int_W^{\sqrt{S}}
{\mathrm{d}}w \frac{{\mathrm{d}}\sigma_k
({\mathrm{e}}^+{\mathrm{e}}^-
\rightarrow b\overline{b})
}{{\mathrm{d}}w},
\label{integ}
\ee
which quantify how much of a given contribution is
located in the region up to a given $W$ ($F_k(W)$) or above it ($G_k(W)$).
As the available data are not copious enough to measure the differential
distribution ${\mathrm{d}}\sigma
({\mathrm{e}}^+{\mathrm{e}}^-
\rightarrow b\overline{b})
/{\mathrm{d}}W$ the theoretical analysis of the distributions (\ref{integ})
might allow us to invent a strategy how to separate the kinematic region of
accessible $W$ into two parts, each dominated by a particular contribution.
The relative importance of the individual contributions as a function
of $W$ is determined by the ratia
\be
r_k(W)\equiv \left.\frac{{\mathrm{d}}\sigma_k({\mathrm{e}}^+{\mathrm{e}}^-
\rightarrow b\overline{b})}{{\mathrm{d}}W}\right/
\frac{{\mathrm{d}}\sigma_{tot}({\mathrm{e}}^+{\mathrm{e}}^-
\rightarrow b\overline{b})}{{\mathrm{d}}W}.
\label{rat}
\ee

\subsection{QED contribution}
\label{subsec:QED}
The pure QED contribution to
$\sigma_{tot}({\mathrm{e}}^+{\mathrm{e}}^-\rightarrow b\overline{b})$
is given as
\be
\frac{{\mathrm{d}}\sigma_{\mathrm{QED}}
({\mathrm{e}}^+{\mathrm{e}}^-\rightarrow b\overline{b})}
{{\mathrm{d}}W}=
\frac{6\alpha^4e_b^4}{\pi S}\frac{A(W)}{W}
\left[\left(1+\frac{4m_b^2}{W^2}-\frac{8m_b^4}{W^4}\right)
\ln\frac{1+\beta}{1-\beta}-\beta\left(1+\frac{4m_b^2}{W^2}\right)\right],
\label{dsigmadwqed}
\ee
where
\be
A(W)=\!\!\!\int\!\!\!\int\!
{\mathrm{d}}y{\mathrm{d}}z
\delta\left(\frac{W^2}{S}-yz\right)
\left[\frac{1+(1-y)^2}{y}\right]
\left[\frac{1+(1-z)^2}{z}\right]
\ln\frac{Q^2_{max}(1-y)}{m_e^2y^2}
\ln\frac{Q^2_{max}(1-z)}{m_e^2z^2},
\label{ay}
\ee
results from convolution of of photon fluxes (\ref{fluxT}), integrated over
the virtualities up to $Q^2_{max}$. The convolution (\ref{ay}) can easily be
performed analytically and the result inserted into (\ref{dsigmadwqed}). In
Fig. \ref{sgamma1}
we display by the solid curve the result of evaluating (\ref{dsigmadwqed}) for
$m_b=4.75$ GeV, $\sqrt{S}=200$ GeV and $Q^2_{max}=4$ GeV. The distribution
vanishes at the threshold $W=2m_b$ due to the threshold behaviour of the
cross section (\ref{pureqed}), peaks at about $W=12$ GeV and then drops
rapidly off due to the fast decrease of both the photon flux (\ref{fluxT})
and (\ref{pureqed}).
\FIGURE{
\epsfig{file=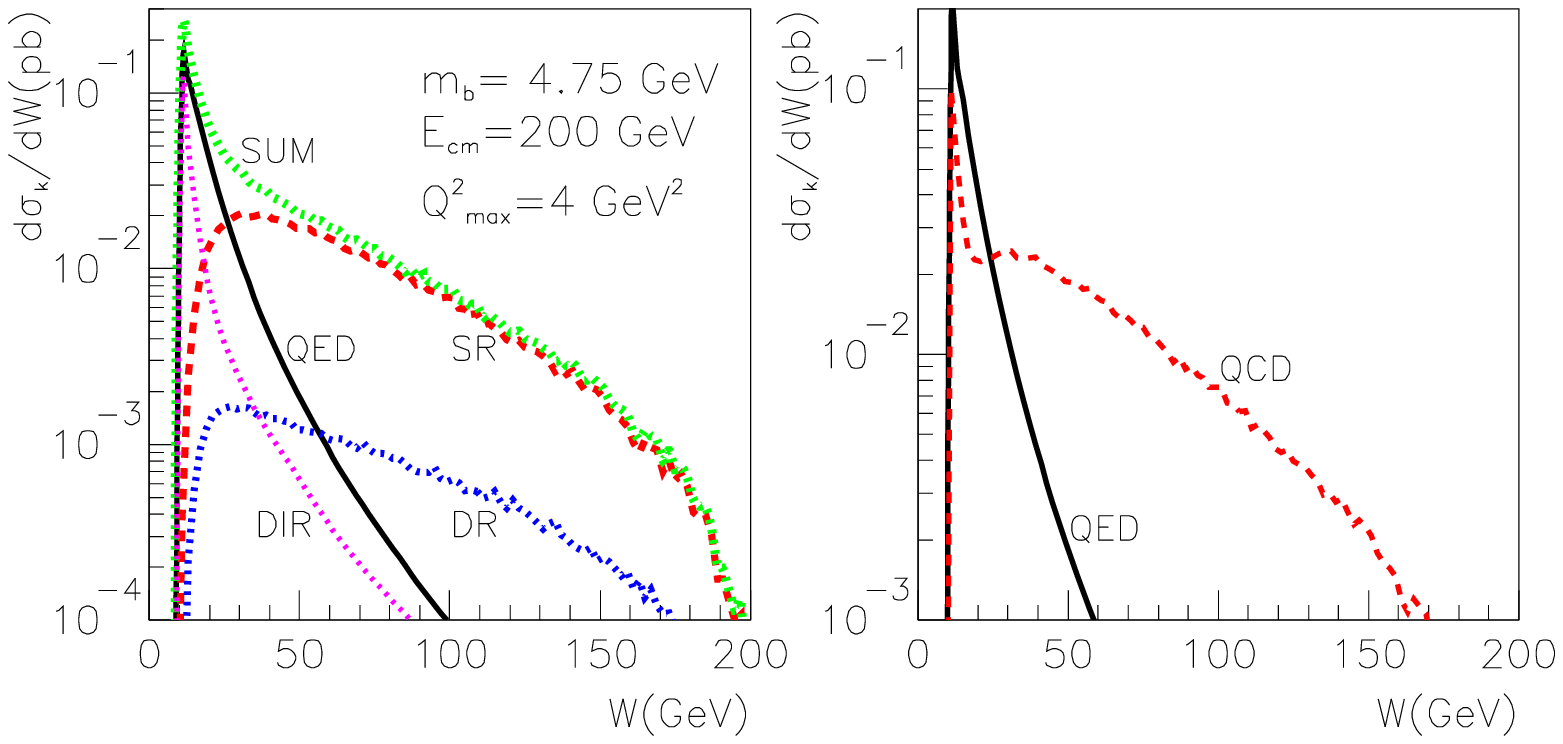,width=14cm}
\caption{Left: The distributions ${\mathrm{d}}\sigma_k/{\mathrm{d}}W$
corresponding to the pure QED contribution together with three
lowest order QCD contributions: single resolved (SR), double resolved (DR),
and direct (DIR). The sum of all four contributions is shown as the upper
dotted curve. Right: the comparison of pure QED contribution with the sum of
three lowest order QCD ones. All curves were obtained for $m_b=4.75$ GeV,
$\sqrt{S}=200$ GeV and $Q^2_{max}=4$ GeV$^2$ using GRV LO PDF of the photon
and setting $\Lambda^{(4)}=0.27$ GeV.}
\label{sgamma1}}
Integrating the distributions in Fig. \ref{sgamma1} yields
the values in the fourth column of Table \ref{table1}.
\begin{table}\centering
\begin{tabular}{|lll|l|l|l|l|l|l|} \hline
\multicolumn{3}{|c|}{Parameters}  & QED  & \multicolumn{3}{c|}{LO QCD} & Total\\
  \hline
   $m_b$ & $\Lambda^{(4)}$ & PDF   &  & DIR  & SR & DR & \\ \hline
$4.75$ & $0.27$ & GRV LO & 1.27 & 0.473 & 1.415 & 0.121 & 3.28\\ \hline
$4.5$ & $0.27$ & GRV LO  & 1.40 & 0.478 & 1.746 & 0.146 & 3.77\\ \hline
$4.75$ & $0.35$ & GRV LO  & 1.27 & 0.520 & 1.542 & 0.141 & 3.47\\ \hline
$4.75$ & $0.27$ & SAS1D  & 1.27 & 0.473 & 0.904 & 0.077 & 2.73 \\ \hline
\end{tabular}
\caption{The integrated cross sections
$\sigma({\mathrm{e}}^+{\mathrm{e}}^-\rightarrow b\overline{b},S)$ for
$\sqrt{S}=200$ GeV and $Q^2_{max}=4$ GeV$^2$, corresponding to the
distributions in Fig. \ref{sgamma1}. The renormalization and factorization
scales $\mu$ and $M$ we identified and set equal to $m_b$. LO form of
$\alpha_s(\mu)$ was used. All cross sections are in picobarns.}
\label{table1}
\end{table}

\subsection{Leading order QCD corrections}
\label{subsec:QCD}
QCD corrections to pure QED expression (\ref{pureqed}) are
of three types: direct (dir), single resolved (sr) and double resolved
(dr). We shall first
discuss the lowest order contributions to all three types of QCD
corrections. As in the case of pure QED contribution, these corrections
are given as convolutions of the photon flux (\ref{fluxT}) with the
appropriate partonic cross sections. In all calculations
$u,d,s$ and $c$ quarks were considered as intrinsic in the photon and
$n_f=4$ was taken in the expression for $\alpha_s(\mu)$.

\subsubsection{Direct photon contribution}
\label{directres}
The $W$ dependence of the leading order QCD correction is given as the
product
\be
\frac{{\mathrm{d}}\sigma_{\mathrm{dir}}^{{\mathrm{LO}}}(W)}
{{\mathrm{d}}W}=
\frac{6\alpha^4e_b^4}{\pi S} \frac{A(W)}{W}\alpha_s(\mu)
\sigma_{\mathrm{dir}}^{(1)}(W/m_b)
\label{LOQCD}
\ee
of the convolution $A(W)$ of photon fluxes and the lowest 
order QCD contribution
$\alpha_s(\mu)\sigma_{\mathrm{dir}}^{(1)}(W)$. At this order the direct
photon contribution $\sigma_{\mathrm{dir}}^{(1)}\alpha_s(\mu)$,
which comes form real or virtual emission of one gluon, is exclusively of
class A. The function $\sigma_{\mathrm{dir}}^{(1)}(W/m_b)$ has been calculated
in, for instance, \cite{johann}. As it is just the first term in the series
in positive powers of $\alpha_s(\mu)$, the value of the
renormalization scale $\mu$ in the argument of $\alpha_s(\mu)$ is
completely arbitrary. The resulting $W$-dependence, evaluated for $\mu=m_b$
and shown in Fig. \ref{sgamma1}, is peaked even more
sharply at small $W$ than the pure QED contribution (\ref{dsigmadwqed}).
This reflects the fact that the cross section
$\sigma_{\mathrm{dir}}^{(1)}(W/m_b)$ does not vanish at the
threshold $W=2m_b$ as does $\sigma_{\mathrm{dir}}^{(0)}(W/m_b)$.

\subsubsection{Resolved photon contribution}
\label{subsec:resolved}
The leading-order single and double resolved photon contributions,
were computed with HERWIG Monte Carlo event generator, which
implements the appropriate LO cross sections of the processes
\begin{eqnarray}
\gamma +G& \rightarrow & b+\overline{b},\label{singler}\\
G+G& \rightarrow & b+\overline{b},
~~q+\overline{q}\rightarrow b+\overline{b},
\label{doubler}
\end{eqnarray}
where $q=u,d,s,c$ stand for intrinsic quarks in the photon,
and convolutes them with photon fluxes and PDF of the quasireal
photon(s). In HERWIG the renormalization
and factorization scales $\mu$ and $M$ are identified and set equal to
an expression which is approximately equal the transverse mass
$\mu=M=M_T\equiv \sqrt{E_T^2+M^2}$. In LEP2 energy range the mean
$\langle\! M_T\!\rangle$ depends weakly on $W$ with, approximately,
$\langle\! M_T\!\rangle\simeq 7$ GeV.

Results of the calculations in which the LO GRV PDF of the photon,
the LO expression for $\alpha_s(\mu)$
with $\Lambda^{(4)}=0.27$ GeV and $m_b=4.75$ GeV were used, are shown in
Fig. \ref{sgamma1}. As expected, the corresponding distributions are much
broader than those of pure QED or LO QCD direct contributions.

\subsection{Comparison of individual contributions}
\label{comparison}
The comparison of the distributions
${\mathrm{d}}\sigma_k/{\mathrm{d}}W, F_k(W)$ and $G_k(W)$, corresponding
to four individual contributions, displayed in Figs. \ref{sgamma1} and
\ref{sgamma2} and summarized in Table \ref{table1}, reveals large
difference in their shapes and magnitude. Specifically we conclude that
\FIGURE{
\epsfig{file=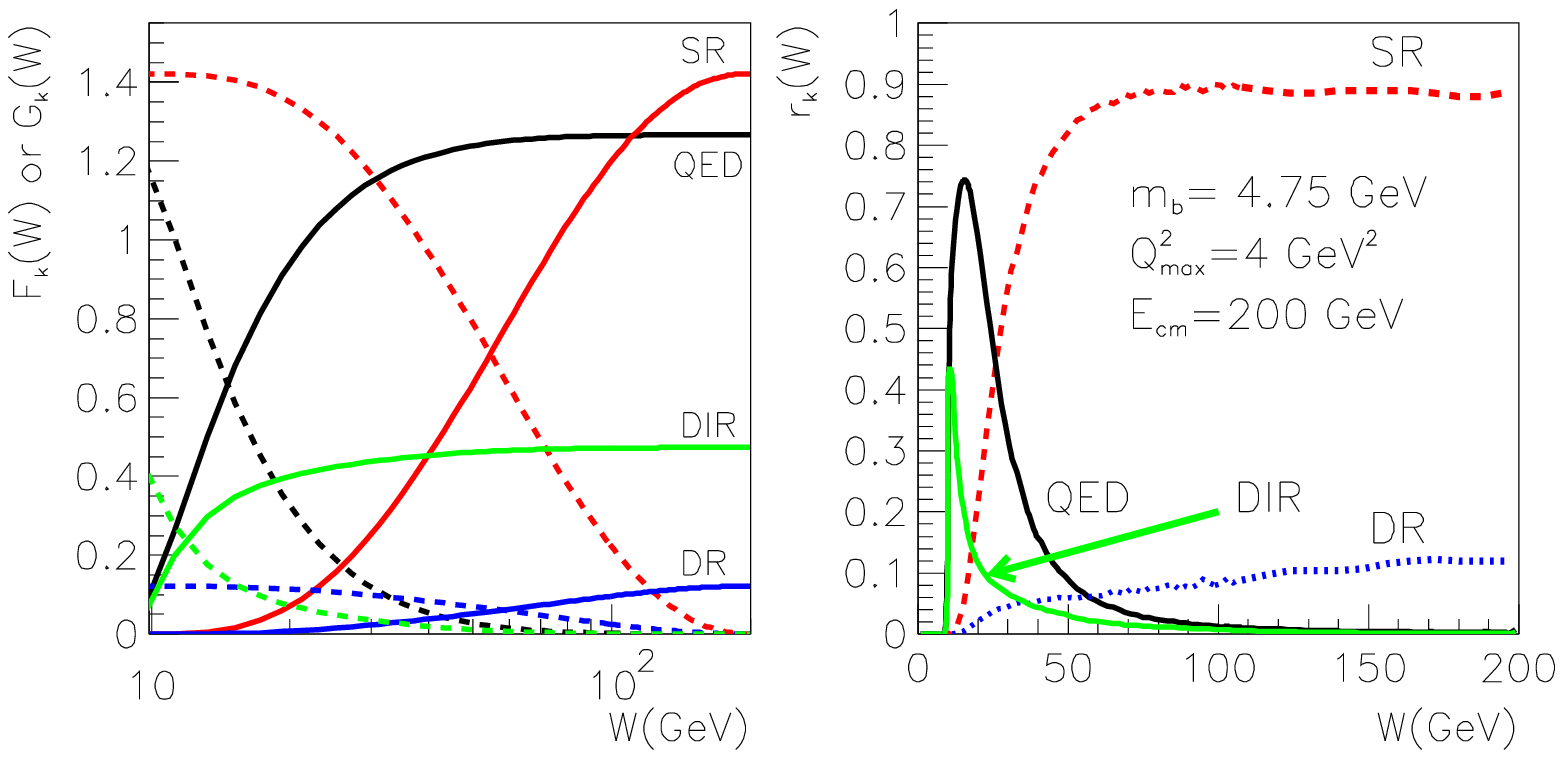,width=14cm}
\caption{Left: solid (dashed) curves show the partially integrated
cross sections $F_k(W)$ ($G_k(W)$) defined in (\ref{integ}) for
QED and three LO QCD contributions. Right: The relative contributions
$r_k(W)$ defined in (\ref{rat}) for the same four contributions.}
\label{sgamma2}}
\begin{itemize}
\item The pure QED as well as the LO direct photon contributions peak at
very small $W$ and are basically negligible above $W\simeq 50$ GeV.
For instance, the left plot of Fig. \ref{sgamma2} shows that $95$\%
of the QED contribution comes from the region $W\lesssim 30$ GeV.
\item The onset of single as well as double resolved
photon contributions is much slower, but these distributions are,
on the other hand, markedly broader.
\item The double resolved photon contribution is negligible
everywhere.
\item The pure QED and single resolved photon contributions are
of comparable size and together make up about 85\% of the total
integrated cross section,
\item up to about $W\simeq 30$ GeV,
${\mathrm{d}}\sigma_{tot}/{\mathrm{d}}W$
is dominated by pure QED contribution, whereas for $W\gtrsim 30$ GeV,
QCD contributions take over.
\end{itemize}
The numbers given in Table 1 correspond to standard colored quarks with
fractional electric charges. In \cite{integer} the excess of data over
standard theoretical calculations is interpreted as evidence for
Hahn-Nambu
integer quark charges. Applied to the case of b-quark, the author of
\cite{integer} argues that the correct way of calculating the charge
factor in (\ref{dsigmadwqed}) is not the usual $3e^4_b=1/27$, but
\be
\frac{1}{3}\left(\sum_{i=1}^{3}e_{b}^{(i)}\right)^4=
\frac{1}{3}=9\frac{1}{27},
\label{fer}
\ee
where the sum runs over the three Hahn-Nambu integer b-quark
charges $e_b^{(i)}$, which are $0,0,-1$ respectively. The results
is thus $9$ times more than in the standard calculation. I think his
argument for first summing in the formula (\ref{fer}) over the
quark colours and then taking the fourth power is wrong
\footnote{The correct value of the charge factor in
(\ref{dsigmadwqed}) for the Hahn-Nambu integer charge b-quark
equals $1$ and would thus yield even bigger enhancement than that
suggested in \cite{integer}.},
but I mention it here because were the author of \cite{integer} right,
the whole discrepancy would come from the region of small $W$,
where QED contribution dominates.

On the other hand, were the light gluino production \cite{Berger}
responsible for the observed excess, the latter would have to come
from the region of $W$ dominated
by the double resolved photon contribution. Although the energy
dependence of the gluon-gluon fusion to gluino-antigluino may be
slightly different than those of $G+G\rightarrow Q\overline{Q}$ or
$q\overline{q}\rightarrow Q\overline{Q}$, it is clear that the basic
shape of the $W$-distribution is given by the convolution of  the
photon fluxes (\ref{fluxT}-\ref{fluxL}) and the gluon distribution
function of the photon, which are the same in both types of processes.

The above observations underline the fact that in order to pin down
the possible origins of the excess of the integrated cross section
$\sigma_{tot}({\mathrm{e}}^+{\mathrm{e}}^-\rightarrow
{\mathrm{e}}^+{\mathrm{e}}^- b\overline{b})$ over theoretical
calculations, it would be very helpful if the data could be separated
at least into two subsamples according to their hadronic energy $W$,
say $W\lesssim 30$ GeV and $W\gtrsim 30$ GeV.

\subsection{Dependence on input parameters}
\label{input} The magnitude of the contributions discussed in the
preceding subsection depend, beside the e$^+$e$^-$ cms energy
$\sqrt{S}$, on a number of input parameters: the numerical values
of $m_b,\Lambda^{(4)}_{\mathrm{QCD}},Q_{max}^2$, the selection of
PDF and the choice of the renormalization and factorization scales
$\mu$ and $M$. In all the calculation reported above we set
$\mu=M=m_b$. The central calculation was performed for
$\sqrt{S}=200$ GeV, $Q^2_{max}=4$ GeV$^2$, $m_b=4.75$ GeV,
$\Lambda^{(4)}=0.27$ GeV using the GRV LO PDF of the photon. To
see the sensitivity of the LO results to these assumptions we
varied some of these parameters:
\begin{itemize}
\item $m_b$ was lowered to $m_b=4.5$ GeV,
\item $\Lambda^{(4)}$ was increased to $0.35$ GeV,
\item GRV set of PDF of the photon was replaced with that of
Schuler-Sj\"{o}strand set SAS1D.
\end{itemize}
The choice of $Q^2_{max}=4$ GeV$^2$ corresponds roughly to
the usual cuts imposed on the LEP2 data and could therefore be also
adjusted to specific conditions of a given experiment.

The results of the calculations of
$\sigma_{tot}({\mathrm{e}}^+{\mathrm{e}}^-\rightarrow
{\mathrm{e}}^+{\mathrm{e}}^- b\overline{b})$,
corresponding to different sets of input parameters specified
above, are listed in Table \ref{table1}. Lowering $m_b$ increases
all four contributions, as does, except for the pure QED one,
increasing $\Lambda^{(4)}$. SAS1D PDF yield
markedly lower results for single and double resolved photon
contributions. It is clear that varying the input parameters
within reasonable bounds does not bring the sum of lowest order QED
and QCD calculations significantly closer to the data.

\section{Can the next-to-leading order QCD corrections solve the
puzzle?}
\label{nlocorrections}
With the sum of lowest order QED and QCD contributions to
$\sigma_{\mathrm{dir}}^{\mathrm{LO}}
({\mathrm{e}}^+{\mathrm{e}}^-\rightarrow
{\mathrm{e}}^+{\mathrm{e}}^-b\overline{b})$
way below the data one might expect that the NLO contributions
could at least partly bridge the gap between data a theory.

\subsection{Direct photon contribution}
The the sum of the second and third terms in (\ref{dir}) can be written,
suppressing the dependence of $\sigma_{\mathrm{dir}}^{(i)}$ on the
ratio $W/m_b$, as
\be
\sigma_{\mathrm{dir}}^{\mathrm{NLO}}=
\sigma_{\mathrm{dir}}^{(1)}\alpha_s(\mu)\left[1+
\frac{\sigma_{\mathrm{dir}}^{(2)}(\mu/m_b)}
{\sigma_{\mathrm{dir}}^{(1)}}\alpha_s(\mu)\right]=
\sigma_{\mathrm{dir}}^{(1)}\alpha_s(\mu)
\left[1+r_1(\mu/m_b)\alpha_s(\mu)\right].
\label{dirNLO}
\ee
Note that $W^2\sigma_{\mathrm{dir}}^{(1)}$ is a unique function of
the ratio $W/m_b$ and the NLO coefficient $r_1(W,m_b,\mu)$ can be
written as a function of $W/m_b$ and $m_b/\mu$. The first term in
(\ref{dirNLO}) is a monotonous function of the renormalization
scale $\mu$, spanning the whole interval between zero and infinity.
As emphasized in Section \ref{dircont}, one needs to include at least
the term $\alpha_s^2\sigma_{\mathrm{dir}}^{(2)}$ to make the expression
(\ref{dirNLO}) of genuine next-to-leading order. Only the class A
of order $\alpha^2\alpha_s^2$ direct photon contributions is needed
for this purpose.  The renormalization scale invariance of
$\sigma_{\mathrm{dir}}^{\mathrm{QCD}}$ implies the following general
form of $r_1$:
\be
r_1(W/m_b,\mu/m_b,{\mathrm{RS}})=\frac{\beta_0}{4\pi}
\ln\frac{\mu^2}{\Lambda^2_{\mathrm{RS}}}
-\rho(W/m_b),
\label{r1}
\ee
where $\rho(W/m_b)$ is a renormalization scale and scheme invariant
\cite{pms}, which, however, depends beside the ratio $W/m_b$ also
on the numerical value of the ratio $m_b/\Lambda^{(4)}$. It can be
evaluated using the results of a calculation in any renormalization
scheme RS
\be
\rho(W/m_b,m_b/\Lambda^{(4)})=
\frac{\beta_0}{4\pi}\ln\frac{m_b^2}{\Lambda^{(4)}_{\mathrm{RS}}}
-r_1(W/m_b,1,{\mathrm{RS}})
\label{rho}
\ee
and its numerical value governs basic features of the scale dependence
of (\ref{dirNLO}):
\FIGURE{
\epsfig{file=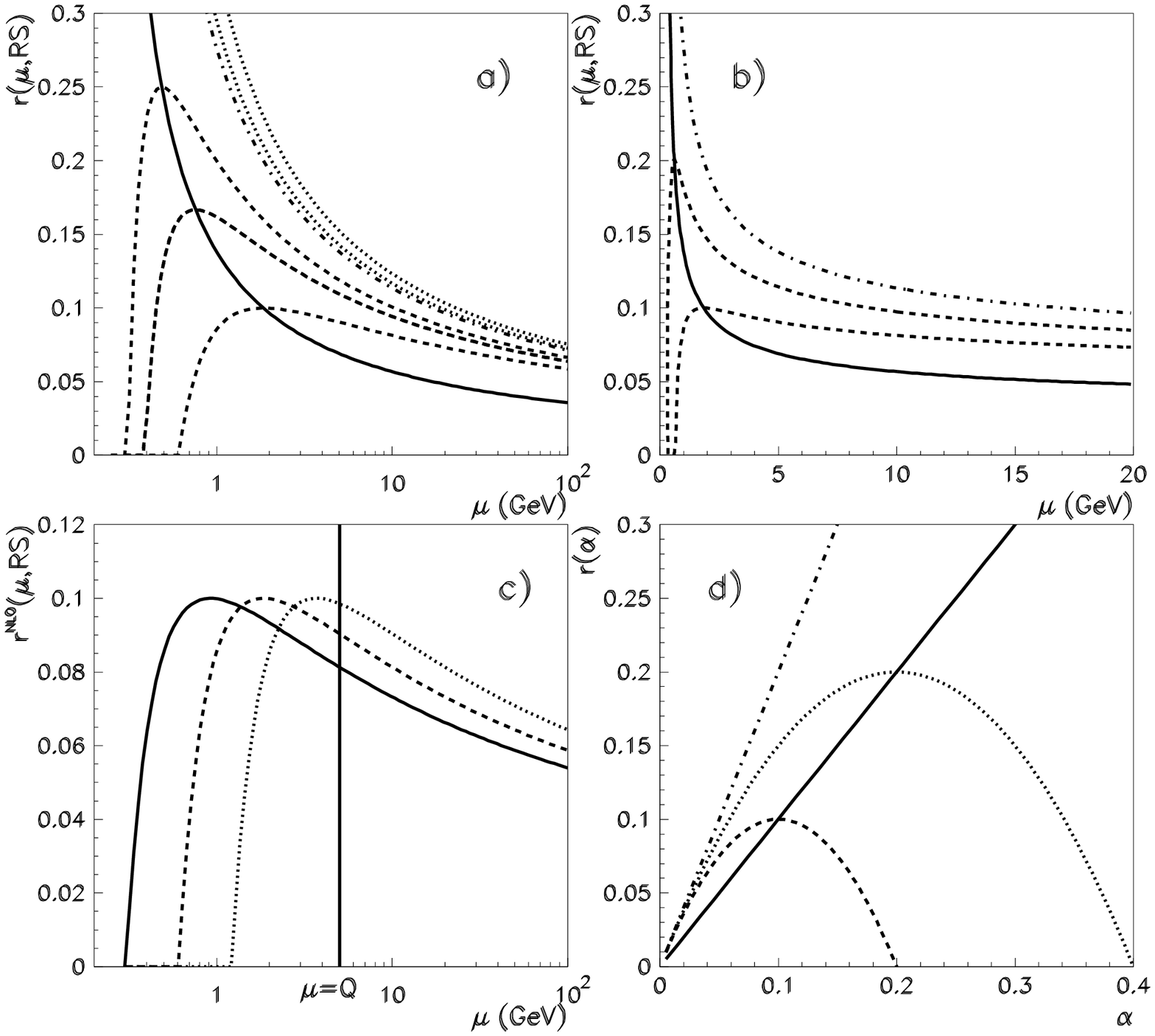,width=12cm}
\caption{a) The renormalization scale dependence of the leading
(solid curve) and next-to-leading order contributions to the
generic quantity (\ref{dirNLOa}) for different values of $\rho$.
The dashed curves correspond to $\rho >0$, the dotted ones to $\rho<0$,
the dash-dotted to $\rho=0$; b) the same as in a) but plotted in a
linear scale of $\mu$; c) graphical representation of (\ref{dirNLOa})
in three different renormalazation schemes and for $\rho >0$; d) the
shape of the NLO expression (\ref{dirNLOa}) as a function of $\alpha_s$.}
\label{figdir2}}
\begin{description}
\item{$\rho>0$}: the NLO approximation (\ref{dirNLO}) considered
as a function of $\mu$ exhibits a local maximum, where
${\mathrm{d}}\sigma_{\mathrm{dir}}^
{\mathrm{NLO}}/{\mathrm{d}}\mu=0$ and where the prediction is thus
most stable. This point, preferred by the Principle of Minimal Sensitivity
\cite{pms}, is also very close to the point for which $r_1=0$,
which is selected by method of Effective Charges \cite{ech}.
The value of $\sigma_{\mathrm{dir}}^{\mathrm{NLO}}$
at this point is proportional to $1/\rho$ implying very large
NLO corrections for small $\rho$. Inserting the appropriate
numbers for $n_f=4$, $m_b=4.75$ GeV and
$\Lambda^{(4)}_{\overline{\mathrm{MS}}}=0.27$ GeV, we get
\be
\rho(W/m_b)=3.88-r_1(W/m_b,1,\overline{\mathrm{MS}}).
\label{rhovalue}
\ee
The NLO coefficient $r_1(W/m_b,1,\overline{\mathrm{MS}})$ does not
therefore have to be outrageously large to get small, or even negative
$\rho$!
\item{$\rho\le 0$:}
$\sigma_{\mathrm{dir}}^{\mathrm{NLO}}$ becomes a monotonous function
of $\mu$, similarly to $\sigma_{\mathrm{dir}}^{\mathrm{LO}}$. In fact,
taking the derivative of $\sigma_{\mathrm{dir}}^{\mathrm{NLO}}$
with respect to $\mu$ one finds that for $\rho<0$ it is
actually even steeper than $\sigma_{\mathrm{dir}}^{\mathrm{LO}}$,
given by the first term in (\ref{dirNLO}). Consequently, for negative
$\rho$ going to the NLO does not improve the stability of the calculation,
but quite on the contrary!
\end{description}
The above features are straightforward to see
setting $\beta_1=0$. This technical assumption simplifies the relevant
formulae significantly, but nothing essential depends on it.
Setting $c=0$ in (\ref{equation}) allows us to write $\alpha_s(\mu)$
explicitly as
\be
\alpha_s(\mu)=\frac{4\pi}{\beta_0 \ln(\mu^2/\Lambda^2_{\mathrm{RS}})}
\label{solution}
\ee
which, inserting this expression into (\ref{dirNLO}) and taking into
account (\ref{r1}), gives
\be
\sigma_{\mathrm{dir}}^{\mathrm{NLO}}=
\sigma_{\mathrm{dir}}^{(1)}\alpha_s(\mu)\left[
2-\rho \alpha_s(\mu)\right],
\label{dirNLOa}
\ee
In Fig. \ref{figdir2} we plot the dependence of the generic
NLO quantity (\ref{dirNLOa}) for several
values, positive as well as negative, of $\rho$ and in different
renormalization schemes. Several conclusions can be drawn from
this figure:
\footnote{These conclusions are well know in the context of perturbative
quantities of the form (\ref{dirNLO}) depending on the renormalization
scale only. We recall them because they will be used in the
analysis of the resolved photon contribution, where the interplay
between the renormalization and factorization scales complicates the
situation.}
\begin{itemize}
\item For $\rho<0$
$\sigma_{\mathrm{dir}}^{\mathrm{NLO}}$, displayed in Fig.
\ref{figdir2}a, is a steeper function of $\mu$ than
$\sigma_{\mathrm{dir}}^{\mathrm{LO}}$ because the second term
in (\ref{dirNLO}), equal to $1-\rho\alpha_s(\mu)$ is monotonously
decreasing function of $\mu$. For positive $\rho$, on the other
hand, (\ref{dirNLOa}) exhibits a local maximum at
$\alpha_s^{max}=1/\rho$.
\item For negative as well as positive values of $\rho$,
$\sigma_{\mathrm{dir}}^{\mathrm{NLO}}\propto 1/\ln \mu$ as
$\mu\rightarrow \infty$. For negative $\rho$
this implies that there is no region of local stability except
for the trivial one at $\mu=\infty$. However, when plotted on a
linear scale of $\mu$ in a limited interval the weak logarithmic
does, as illustrated in Fig. \ref{figdir2}b, fake the local
quasistability.
\item The curve representing $\sigma_{\mathrm{dir}}^{\mathrm{NLO}}$
depends, as shown in Fig. \ref{figdir2}c for $\rho >0$, on the
chosen renormalization scheme. Setting $\mu$ equal to some
``natural'' physical scale $Q$ therefore does not resolve the
renormalization scale ambiguity as in different RS we get
different results.
However, although also the position of the
local maximum depends on the choice of the RS, the value of the
NLO approximation (\ref{dirNLOa}) does not! The same holds for
the intersection of the LO and NLO curves.
\item Instead of varying both the renormalization scale
and scheme, which is legitimate but redundant, we may use
the couplant $\alpha_s$ itself for labeling the different
predictions of (\ref{dirNLOa}). Instead of an infinite set
of curves describing the $\mu$-dependence of
$\sigma_{\mathrm{dir}}^{\mathrm{NLO}}(\mu,{\mathrm{RS}})$
in different RS, we get for each $\rho$ a single curve
displayed in Fig. \ref{figdir2}d.
\end{itemize}
We wish to emphasize that whereas there are natural
physical scales in hard collisions, there is nothing like the
``natural'' renormalization scheme. There are only two general
lines of arguments for choosing the renormalization scheme:
either one looks for the maximum local stability or smallest
order $\alpha_s^2$ corrections. In the first case one is led to
the Principle of Maximal Stability \cite{pms}, in the second
to the method of Effective Charges \cite{ech}. As illustrated
by the situation for negative $\rho$, there may, however, he
cases when none of them can be applied.

As the term $\sigma_{\mathrm{dir}}^{(2)}$ in (\ref{dirNLO}) has
not yet been calculated, we cannot associate the class A direct
photon contribution, given by the first term in (\ref{dirNLO}), to a
well-defined renormalization scheme. Moreover, as the magnitude of
the NLO corrections in (\ref{dirNLO}) is determined by the ratio
$\sigma_{\mathrm{dir}}^{(2)}(W/m_b,1)/
\sigma_{\mathrm{dir}}^{(1)}(W/m_b)$ of two functions of $W/m_b$,
which may depend on $W/m_b$ in different ways, the coefficient
$r_1$ may be very large even when both the numerator and denominator
are on average of comparable and small magnitude. The size of
NLO correction may also depend sensitively on the ratio $W/m_b$.
All this indicates that without the knowledge of class A
direct photon contribution of the order $\alpha^2\alpha_s^2$, we cannot
make a meaningful estimate of the importance of higher order
corrections to (\ref{dirNLO}).

\subsection{Resolved photon contribution}
As in LEP2 energy range the double resolved photon one is numerically
negligible, only the single resolved photon contribution will discussed in
detail below. As shown in Fig. \ref{sgamma1}, the spectrum of
the contributions ${\mathrm{d}}\sigma_{\mathrm{sr}}/{\mathrm{d}}W$ peaks at
about $W=30$ GeV with the mean value $\langle W\rangle \doteq 65$ GeV. The
properties of the measures cross section
$\sigma_{\mathrm{dir}}^{\mathrm{NLO}}
({\mathrm{e}}^+{\mathrm{e}}^-\rightarrow
{\mathrm{e}}^+{\mathrm{e}}^-b\overline{b})$
will therefore be determined primarily by those
of $\sigma_{\mathrm{dir}}^{\mathrm{NLO}}
(\gamma\gamma\rightarrow b\overline{b})$
in the energy range $30\lesssim W\lesssim 65$ GeV.
\FIGURE{
\epsfig{file=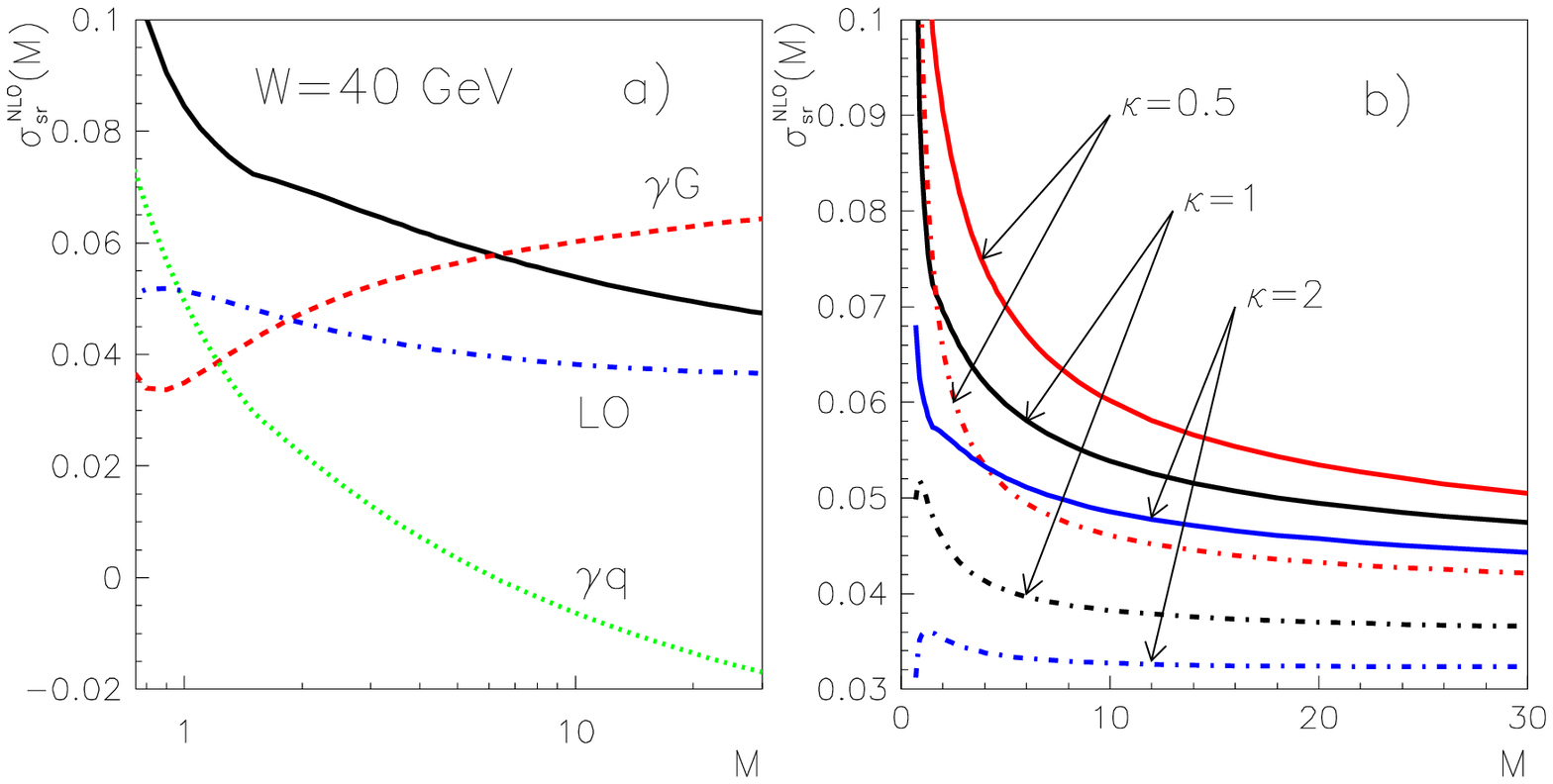,width=12cm}
\caption{a) The scale dependence of the conventional NLO approximation
$\sigma^{\mathrm{NLO}}_{\mathrm{sr}}(W,M,M)$ for $W=40$ GeV (solid curve)
together with the contributions of the $\gamma G$ (dashed) and
$\gamma q/\overline{q}$ (dotted) channels. The LO approximation is shown
for comparison by the dash-dotted curve.
b) $\sigma^{\mathrm{NLO}}_{\mathrm{sr}}(W,M,\kappa M)$ (solid curve) and
$\sigma^{\mathrm{LO}}_{\mathrm{sr}}(W,M,\kappa M)$ (dash-dotted) for $W=40$
GeV and three values of $\kappa=0.5,1.2$.}
\label{figsres1}}
All the results shown below for the single resolved photon contribution
\begin{eqnarray}
\sigma^{\mathrm{NLO}}_{\mathrm{sr}}(W,M,\mu)&=&
2\alpha_s(\mu)\int{\mathrm{d}}xG(x,M)
\left[\sigma_{\gamma G}^{(1)}(x)+\alpha_s(\mu)\sigma_{\gamma G}^{(2)}
(x,M,\mu)\right]+\label{l34}\nonumber \\
& & 4\alpha_s^2(\mu)\int{\mathrm{d}}x\sum_iq_i(x,M)
\sigma_{\gamma q_i}^{(2)}(x,M)
\label{singleres}
\end{eqnarray}
are based on the formulae for the partonic cross sections
$\sigma^{(k)}_{ij}$ as given in \cite{ellis}. Even if the reader does not
agree with our claim that the approximations employed in
\cite{zerwas,kramer,laenen} do not constitute complete NLO approximation,
it is certainly important to understand quantitatively their
renormalization and factorization scale dependence.
Because the expressions for $\sigma_{\gamma G}^{(2)}$ as given in
\cite{ellis} correspond to $\mu=M$, we have restored its separate
dependence on $\mu$ and $M$ by adding to
$\sigma_{\mathrm{\gamma G}}^{(2)}(x,M,M)$
the term $(\beta_0/4\pi)\sigma_{\gamma G}^{(1)}\ln(\mu^2/M^2)$.
Note that for each value of factorization scale $M$ the
expression (\ref{singleres}) has, as far as the $\mu$-dependence is
concerned, the form of the NLO expression
(\ref{dirNLO}). In all calculations the GRV HO
set of PDF of the photon and $\Lambda_{\overline{\mathrm{MS}}}^{(4)}=0.27$
GeV were used.

We first follow the conventional procedure and set $M=\mu$.
The resulting (common) scale dependence of the expression
(\ref{singleres}), together with those of the quark and gluon
contributions to it, are shown in Fig. \ref{figsres1}a for $W=40$ GeV.
Overlaid for comparison is also the LO approximation, given by the
first term in (\ref{singleres}). We note the different scale
dependence of the $\gamma G$ and $\gamma q$ channels, the latter turning
negative for $M\gtrsim 6$ GeV, but the most important observation
concerns the fact that the conventional NLO approximation (\ref{singleres})
is a monotonously decreasing function of the scale. Moreover, it falls off
even more steeply than the LO expression! In other words in going from the
leading to the next-to-leading order the sensitivity to the scale variation
increases, rather than decreases, as one might expect (and hope)!
\FIGURE{
\epsfig{file=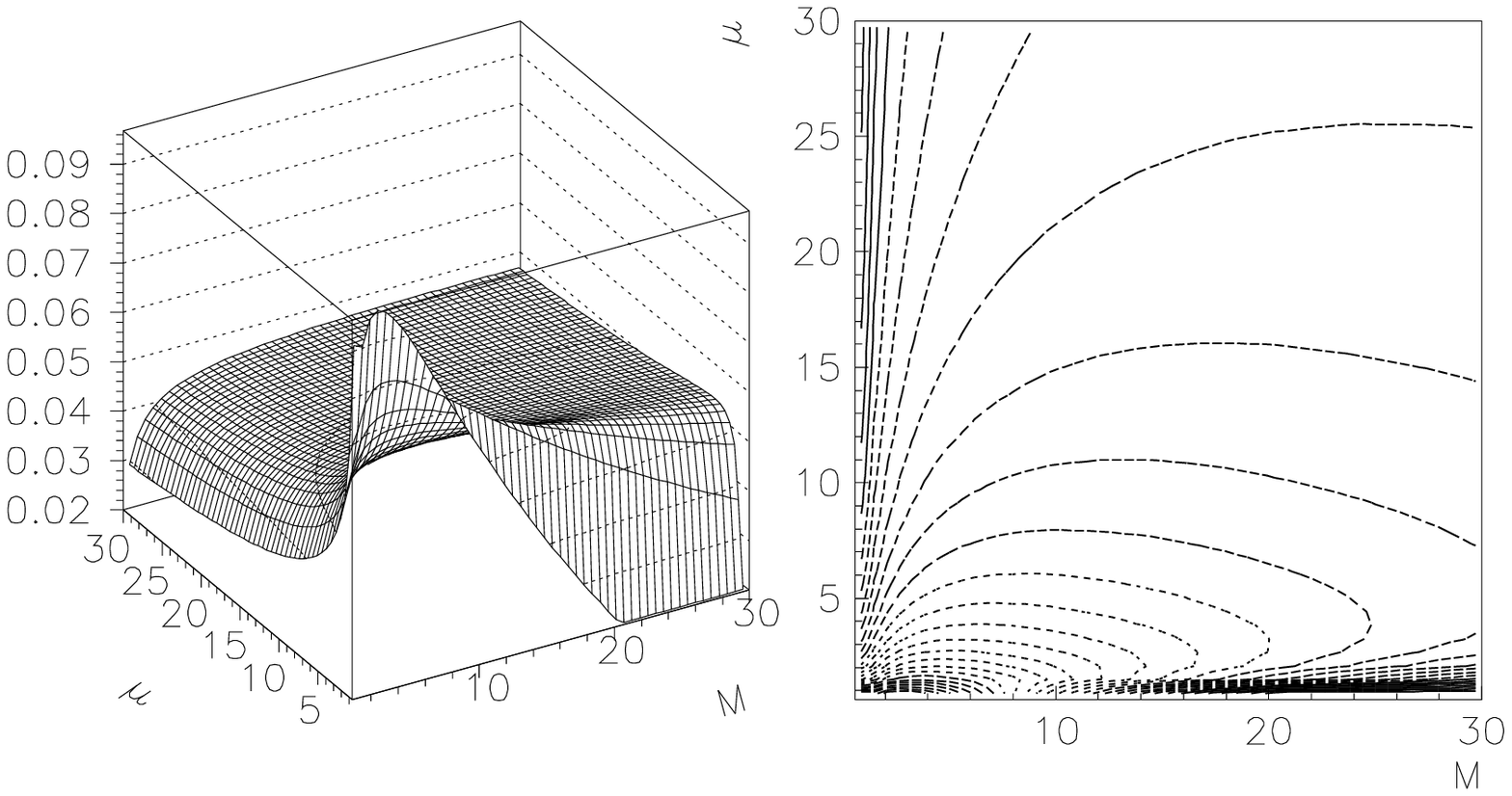,width=12cm}
\caption{The surface and contour plots of
$\sigma^{\mathrm{NLO}}_{\mathrm{sr}}(W,M,\mu)$ for $W=40$ GeV.}
\label{figsres2}}
Recalling the discussion in Section 5.1 one should, however, not be
surprised. To check how much this feature depends on setting exactly
$\mu=M$, we plot in Fig. \ref{figsres1}b the scale dependence of
$\sigma_{\mathrm{sr}}^{\mathrm{NLO}}(W,M,\mu=\kappa M)$
for standard choices of $\kappa=0.5,1,2$. Clearly, the above
conclusion is independent of $\kappa$ in this range.

The steep and monotonous scale dependence of
$\sigma_{\mathrm{sr}}^{\mathrm{NLO}}(W,M,\mu=\kappa M)$
is a clear warning that the conventional NLO approximation
is highly unstable. To see what happens if we
relax the usual but arbitrary identification $\mu=\kappa M$
we plot in Fig. \ref{figsres2} the surface and contour plots
representing the full $M$ and $\mu$ dependence of
$\sigma_{\mathrm{sr}}^{\mathrm{NLO}}(W,M,\mu)$ as given
in eq. (\ref{singleres}).
\FIGURE{
\epsfig{file=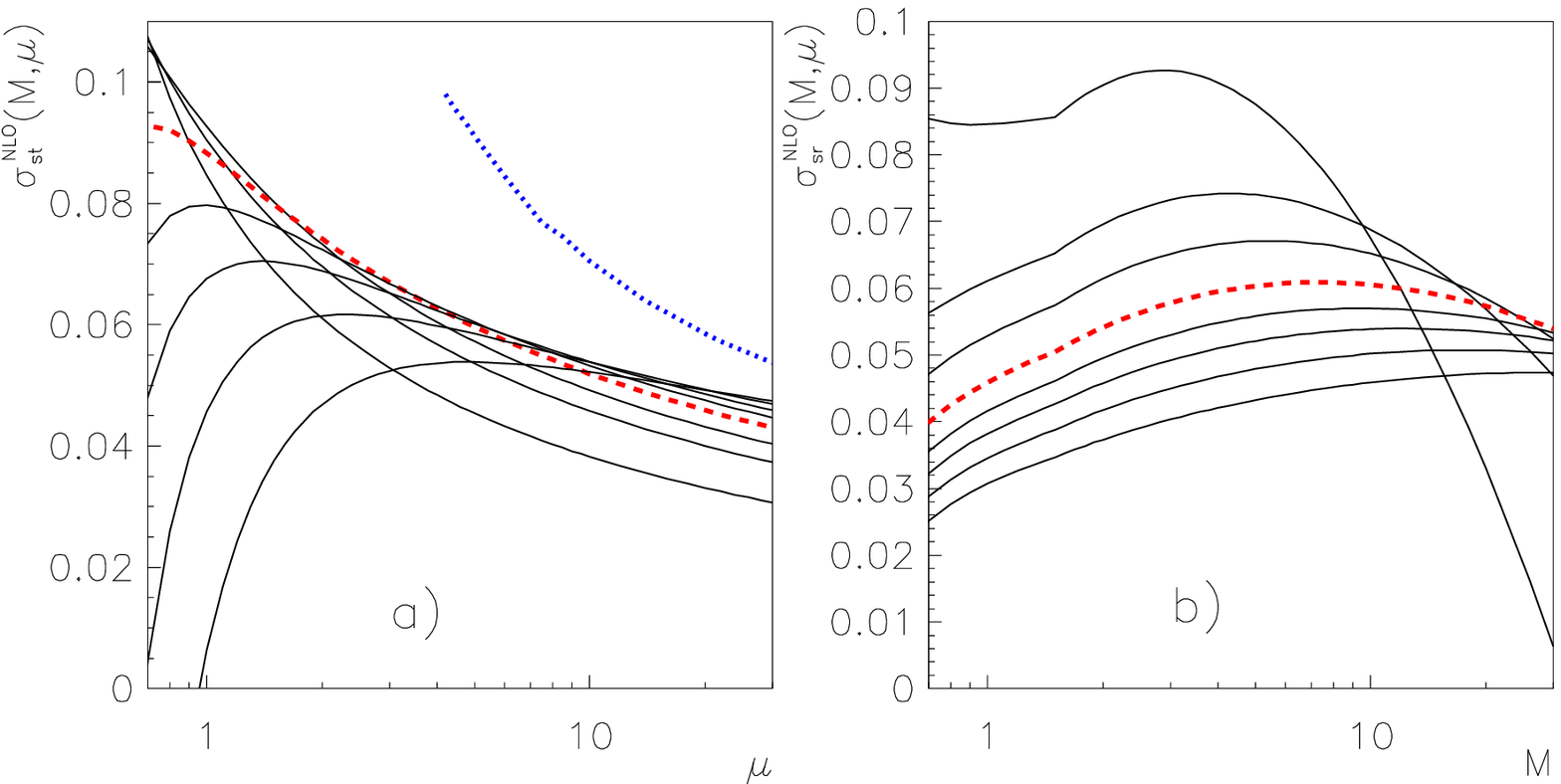,width=12cm}
\caption{The renormalization scale dependence of
$\sigma^{\mathrm{NLO}}_{\mathrm{sr}}(W,M,\mu)$ for fixed values of the
factorization scale (a) and vice versa: the
factorization scale dependence of
$\sigma^{\mathrm{NLO}}_{\mathrm{sr}}(W,M,\mu)$ for
the same set of fixed values of the renormalization scale $\mu$ (b).
All calculations correspond to $W=40$ GeV. In a) the ordering
from above of the curves at $\mu=30$ GeV corresponds to
$M=30,16,10,7,4.75,3,2,1$ GeV, in b) the curves correspond at $M=0.7$
GeV to the same sequence from below.}
\label{figsres3}}
Contrary to analogous process in antiproton-proton collisions \cite{ja},
it does not exhibit a  saddle point, 
where the derivatives with respect to both $M$ and $\mu$ vanish, 
but Fig. \ref{figsres2} seems to indicate some
sort of stability region at large scales, say for $M\gtrsim 10$ GeV,
$\mu\gtrsim 20$ GeV. This impression is, however, misleading
as becomes clear if we plot in Fig. \ref{figsres3} the
slices of the surface plot in Fig. \ref{figsres2}a along
both axis and recall the discussion of Section 5.1. For each
fixed value of the factorization scale $M$ the expression
(\ref{singleres}) has a form of the NLO expression as far as the
renormalization scale $\mu$ is concerned. Comparing the curves
in Fig. \ref{figsres3}a with those of \ref{figdir}a we see that
for $M\lesssim 4.2$ GeV $\sigma^{\mathrm{NLO}}(M,\mu)$ corresponds
to negative $\rho$ in (\ref{dirNLOa}) and thus exhibits no local
stability point. For higher $M$ the local maximum in the
$\mu$-dependence of $\sigma^{\mathrm{NLO}}_{\mathrm{sr}}(M,\mu)$
exists at the associated $\mu_{max}(M)$. The $M$-dependence
of $\sigma^{\mathrm{NLO}}_{\mathrm{sr}}(M,\mu_{max}(M))$, shown in Fig.
\ref{figsres3}a by the dotted curve, is, however, even steeper that those
at fixed $M$. The above plots and conclusions concerned the results at one
typical value of $W$, but their essence holds for the whole
interval relevant for LEP2 data.

We thus conclude that in the energy range relevant for
LEP2 data the renormalization and factorization scale dependence of the
conventional NLO calculations of single resolved
photon contribution to the total cross section
$\sigma_{tot}(\gamma\gamma\rightarrow b\overline{b})$ exhibits no stability
region, either as a function of the common scale $\mu=\kappa M$ or
as fully two dimensional function of $\mu$ and $M$.

\section{Summary and Conclusions}
\label{summary}
We have argued that in order to understand the origins of the excess of LEP2
data on $b\overline{b}$ production in $\gamma\gamma$ collisions over the
current theoretical calculations, two ingredients are needed.

On the experimental side, the separation of data into at least two bins of
the hadronic energy $W$, say $W\lesssim 30$ GeV and $W\gtrsim 30$ GeV,
could be instrumental in pinning down the possible mechanisms or phenomena
responsible for the observed excess.

On the theoretical side, the evaluation of the direct photon contribution
of the order $\alpha^2\alpha_s^2$ is needed to provide the terms that
would make the existing theoretical expressions of genuine next-to-leading
order in $\alpha_s$. In their absence, the existing
NLO calculations are highly sensitive to the variation of renormalization
and factorization scale and thus inherently unreliable.

\section*{Acknowledgments}
The work has been supported by the Ministry of Education of the
Czech Republic under the project LN00A006.

\end{document}